\PassOptionsToPackage{unicode=true}{hyperref} % options for packages loaded elsewhere
\PassOptionsToPackage{hyphens}{url}
\PassOptionsToPackage{dvipsnames,svgnames*,x11names*}{xcolor}
\documentclass[,man,floatsintext]{apa6}
\usepackage{lmodern}
\usepackage{amssymb,amsmath}
\usepackage{ifxetex,ifluatex}
\usepackage{fixltx2e} % provides \textsubscript
\ifnum 0\ifxetex 1\fi\ifluatex 1\fi=0 % if pdftex
  \usepackage[T1]{fontenc}
  \usepackage[utf8]{inputenc}
  \usepackage{textcomp} % provides euro and other symbols
\else % if luatex or xelatex
  \usepackage{unicode-math}
  \defaultfontfeatures{Ligatures=TeX,Scale=MatchLowercase}
\fi
\IfFileExists{upquote.sty}{\usepackage{upquote}}{}
\IfFileExists{microtype.sty}{%
\usepackage[]{microtype}
\UseMicrotypeSet[protrusion]{basicmath} % disable protrusion for tt fonts
}{}
\IfFileExists{parskip.sty}{%
\usepackage{parskip}
}{% else
\setlength{\parindent}{0pt}
\setlength{\parskip}{6pt plus 2pt minus 1pt}
}
\usepackage{xcolor}
\usepackage{hyperref}
\hypersetup{
            pdftitle={Toward a principled Bayesian workflow in cognitive science},
            pdfauthor={Daniel J. Schad, Michael Betancourt, \& Shravan Vasishth},
            pdfkeywords={Workflow, prior predictive checks, posterior predictive checks, model building, Bayesian data analysis},
            colorlinks=true,
            linkcolor=Maroon,
            filecolor=Maroon,
            citecolor=Blue,
            urlcolor=blue,
            breaklinks=true}
\usepackage{color}
\usepackage{fancyvrb}

\DefineVerbatimEnvironment{Highlighting}{Verbatim}{commandchars=\\\{\}}
\usepackage{framed}
\definecolor{shadecolor}{RGB}{248,248,248}
\newenvironment{Shaded}{\begin{snugshade}}{\end{snugshade}}

\newcommand{\DataTypeTok}[1]{\textcolor[rgb]{0.13,0.29,0.53}{#1}}
\newcommand{\DecValTok}[1]{\textcolor[rgb]{0.00,0.00,0.81}{#1}}

\newcommand{\KeywordTok}[1]{\textcolor[rgb]{0.13,0.29,0.53}{\textbf{#1}}}
\newcommand{\NormalTok}[1]{#1}
\newcommand{\OperatorTok}[1]{\textcolor[rgb]{0.81,0.36,0.00}{\textbf{#1}}}
\newcommand{\OtherTok}[1]{\textcolor[rgb]{0.56,0.35,0.01}{#1}}

\newcommand{\StringTok}[1]{\textcolor[rgb]{0.31,0.60,0.02}{#1}}

\usepackage{longtable,booktabs}
\IfFileExists{footnote.sty}{\usepackage{footnote}\makesavenoteenv{longtable}}{}
\usepackage{graphicx,grffile}
\makeatletter
\def\maxwidth{\ifdim\Gin@nat@width>\linewidth\linewidth\else\Gin@nat@width\fi}
\def\maxheight{\ifdim\Gin@nat@height>\textheight\textheight\else\Gin@nat@height\fi}
\makeatother
\setkeys{Gin}{width=\maxwidth,height=\maxheight,keepaspectratio}
\providecommand{\tightlist}{%
  \setlength{\itemsep}{0pt}\setlength{\parskip}{0pt}}
\ifx\paragraph\undefined\else
\let\oldparagraph\paragraph
\renewcommand{\paragraph}[1]{\oldparagraph{#1}\mbox{}}
\fi
\ifx\subparagraph\undefined\else
\let\oldsubparagraph\subparagraph
\renewcommand{\subparagraph}[1]{\oldsubparagraph{#1}\mbox{}}
\fi

\makeatletter
\def\fps@figure{htbp}
\makeatother

\shorttitle{Bayesian workflow for cognitive science}
\affiliation{
\vspace{0.5cm}
\textsuperscript{1} University of Potsdam, Germany\\\textsuperscript{2} Tilburg University, Netherlands\\\textsuperscript{3} Symplectomorphic, New York, USA}
\keywords{Workflow, prior predictive checks, posterior predictive checks, model building, Bayesian data analysis}
\usepackage{csquotes}
\usepackage{upgreek}
\usepackage{longtable}
\usepackage{lscape}
\usepackage{multirow}
\usepackage{tabularx}
\usepackage[flushleft]{threeparttable}
\usepackage{threeparttablex}

\makeatletter
\newcommand\LastLTentrywidth{1em}
\newlength\longtablewidth
\newcommand{\getlongtablewidth}{\begingroup \ifcsname LT@\roman{LT@tables}\endcsname \global\longtablewidth=0pt \renewcommand{\LT@entry}[2]{\global\advance\longtablewidth by ##2\relax\gdef\LastLTentrywidth{##2}}\@nameuse{LT@\roman{LT@tables}} \fi \endgroup}

\makeatletter
\renewcommand{\paragraph}{\@startsection{paragraph}{4}{\parindent}%
  {0\baselineskip \@plus 0.2ex \@minus 0.2ex}%
  {-1em}%
  {\normalfont\normalsize\bfseries\itshape\typesectitle}}

\renewcommand{\subparagraph}[1]{\@startsection{subparagraph}{5}{1em}%
  {0\baselineskip \@plus 0.2ex \@minus 0.2ex}%
  {-\z@\relax}%
  {\normalfont\normalsize\itshape\hspace{\parindent}{#1}\textit{\addperi}}{\relax}}
\makeatother

\usepackage{etoolbox}
\makeatletter
\patchcmd{\HyOrg@maketitle}
  {\section{\normalfont\normalsize\abstractname}}
  {\section*{\normalfont\normalsize\abstractname}}
  {}{\typeout{Failed to patch abstract.}}
\makeatother
\usepackage{amsmath}
\usepackage[american]{babel}
\usepackage[utf8]{inputenc}
\usepackage[T1]{fontenc}
\usepackage[backend=biber,useprefix=true,style=apa,url=false,doi=false,sorting=nyt,eprint=false]{biblatex}
\DeclareLanguageMapping{american}{american-apa}
\usepackage{fancyvrb}
\usepackage{newfloat}
\usepackage{graphicx}
\usepackage{caption}
\usepackage{listings}
\usepackage{placeins}

\title{Toward a principled Bayesian workflow in cognitive science}
\author{Daniel J. Schad\textsuperscript{1, 2}, Michael Betancourt\textsuperscript{3}, \& Shravan Vasishth\textsuperscript{1}}
\date{}

\authornote{The principled Bayesian workflow, with a different (non-cognitive) running example analysis, was previously developed and documented by Betancourt (2018) on \url{http://bit.ly/396lGSX}. Moreover, the content of this manuscript, including the Bayesian workflow and the cognitive example data analysis, was presented by DJS as a talk at the 52nd Annual Meeting of the Society for Mathematical Psychology 2019 in Montreal. The example experimental data analyzed in this manuscript was previously published by Gibson \& Wu (2013).

Correspondence concerning this article should be addressed to Daniel J. Schad, Tilburg University, Netherlands. E-mail: \href{mailto:danieljschad@gmail.com}{\nolinkurl{danieljschad@gmail.com}}}

\abstract{
Experiments in research on memory, language, and in other areas of cognitive science are increasingly being analyzed using Bayesian methods. This has been facilitated by the development of probabilistic programming languages such as Stan, and easily accessible front-end packages such as brms. The utility of Bayesian methods, however, ultimately depends on the relevance of the Bayesian model, in particular whether or not it accurately captures the structure of the data and the data analyst's domain expertise. Even with powerful software, the analyst is responsible for verifying the utility of their model. To demonstrate this point, we introduce a principled Bayesian workflow (Betancourt, 2018) to cognitive science. Using a concrete working example, we describe basic questions one should ask about the model: prior predictive checks, computational faithfulness, model sensitivity, and posterior predictive checks. The running example for demonstrating the workflow is data on reading times with a linguistic manipulation of object versus subject relative clause sentences. This principled Bayesian workflow also demonstrates how to use domain knowledge to inform prior distributions. It provides guidelines and checks for valid data analysis, avoiding overfitting complex models to noise, and capturing relevant data structure in a probabilistic model. Given the increasing use of Bayesian methods, we aim to discuss how these methods can be properly employed to obtain robust answers to scientific questions. All data and code accompanying this paper are available from \url{https://osf.io/b2vx9/}.

}

\begin{document}
\maketitle

{
\hypersetup{linkcolor=}
\setcounter{tocdepth}{3}
\tableofcontents
}
\hypertarget{introduction}{%
\section{Introduction}\label{introduction}}

Recent years have seen a rise in the use of Bayesian statistics for data analysis in the cognitive sciences and other areas.
\textcolor{black}{There are many perspectives on Bayesian inference, especially in cognitive science;}
for recent overviews see special issues \textcolor{black}{in the Journal of Mathematical Psychology} (Lee, 2011; Mulder \& Wagenmakers, 2016), \textcolor{black}{in Psychological Methods} (Chow \& Hoijtink, 2017; Hoijtink \& Chow, 2017), \textcolor{black}{and in Psychonomic Bulletin \& Review} (Vandekerckhove, Rouder, \& Kruschke, 2018); \textcolor{black}{for introductory articles see} Etz \& Vandekerckhove (2018) and Etz et al. (2018).
Th\textcolor{black}{e} rise \textcolor{black}{in the use of Bayesian statistics} has been fueled by increasing recognition of the advantages of robust Bayesian analyses (Gelman et al., 2014).
\textcolor{black}{In this paper we discuss a workflow to help build Bayesian analyses of principled models that strive to capture the relevant details of the processes that generate data and the domain expertise pertinent to those processes.}

\textcolor{black}{For the field of cognitive science, one advantage is that Bayesian analyses are more robust than frequentist equivalents as they regularize inference in low-power situations} (Gelman et al., 2014; Morey, Romeijn, \& Rouder, 2016). \textcolor{black}{Importantly, Bayesian methods provide a possibility to quantify uncertainty about cognitive parameters and models, which is not provided by frequentist approaches} (Wagenmakers, Morey, \& Lee, 2016). \textcolor{black}{Moreover, Bayesian inference works the same for statistical and process models, yielding a general framework for statistical and computational analyses} (Lee, 2011; Lee \& Wagenmakers, 2014). \textcolor{black}{Specifically, nonlinear hierarchical models are conceptually and computationally inconvenient in frequentist contexts, but are conceptually simple and computationally tractable in Bayesian frameworks} (for some examples of Bayesian hierarchical models, see, Morey, 2011; Ly et al., 2017; Nilsson, Rieskamp, \& Wagenmakers, 2011; Pooley, Lee, \& Shankle, 2011; Pratte \& Rouder, 2011)\textcolor{black}{. More generally, the Bayesian framework speeds up the process of developing and fitting hierarchical and mixed models} (Gelman et al., 2014). \textcolor{black}{In these models, the prior is that "people are not so dissimilar", and regularization falls out as a natural consequence. An alternative way to use priors as part of the modeling is to encode different sets of beliefs in the prior } (Haaf \& Rouder, 2017), \textcolor{black}{whereby different theoretical hypotheses are encoded in the prior.}

\textcolor{black}{Most Bayesian posterior estimation requires software for posterior sampling.} Much progress has been made in the development of probabilistic programming languages such as Stan (Carpenter et al., 2017), WinBUGS (Lunn, Thomas, Best, \& Spiegelhalter, 2000), \textcolor{black}{JASP} (JASP Team, 2019), and JAGS (Plummer, 2012). Packages like \emph{brms} (Bürkner, 2017b) \textcolor{black}{and \textit{rstanarm}} (Goodrich, Gabry, Ali, \& Brilleman, 2018) now provide easy access to fitting complex hierarchical (non-)linear mixed models in the R System for Statistical Computing (R Core Team, 2016). Front-ends like \emph{brms} \textcolor{black}{and \textit{rstanarm}} have the advantage that standardly used models in cognitive science can be fit using a familiar syntax that is well-known from fitting frequentist linear mixed effects models (the \emph{lme4} package, Baayen, Davidson, \& Bates, 2008; Bates, Mächler, Bolker, \& Walker, 2015).

Although complex and powerful, Bayesian analysis tools are thus now easily accessible to lay users, and although these tools greatly facilitate Bayesian computations, the model specification is still (as it should be) the responsibility of the user. The steps needed to arrive at a useful and robust analysis, however, are usually not spelt out in introductory textbooks or tutorial articles. The present paper seeks to fill this gap in the literature. \textcolor{black}{It assumes that readers have had some previous exposure to and experience with basic concepts of Bayesian modeling, such as fitting simple Bayesian models (e.g., in brms or Stan)} (Bürkner, 2017a; Kruschke, 2014; Vasishth et al., 2018b) \textcolor{black}{, and are interested in how to put together their knowledge about basic applications into a robust and principled workflow.}

Much research has been carried out in recent years to develop \textcolor{black}{complementary} tools to ensure robust Bayesian data analyses (e.g., Gabry, Simpson, Vehtari, Betancourt, \& Gelman, 2017; Talts, Betancourt, Simpson, Vehtari, \& Gelman, 2018). One of the \textcolor{black}{key features} of this research has been the formulation of a principled Bayesian workflow for conducting a probabilistic analysis (Betancourt, 2018), \textcolor{black}{which emphasizes the interplay between domain expertise encoded in the prior model and information encoded in the likelihood function.}
\textcolor{black}{Note that this prior knowledge is only interpretable in the context of the likelihood} (Gelman, Simpson, \& Betancourt, 2017).
This workflow provides an initial coherent picture of steps to take for a robust analysis (also see e.g., Gabry et al., 2017), leaving room for further improvements and methodological developments. At an abstract level, parts of this workflow can be applied to any kind of data analysis, be it frequentist or Bayesian, be it based on sampling or on analytic procedures\textcolor{black}{, and be it for linear statistical models or for non-linear cognitive process models}.

The papers cited above, however, are written either for a general audience or for the professional statistics researcher. For newcomers to Bayesian methods, translating these ideas to their own domain is often very difficult to impossible. What is needed is an explicit, reproducible, fully worked-out example of the workflow for a common type of experimental design. The main challenge with field-specific translations of statistical methods is that they need to be accessible to a non-technical audience but at the same time uncompromising on the details. \textcolor{black}{Thus, the principled workflow we present is general to any kind of modeling. The present paper} \textcolor{black}{demonstrates} \textcolor{black}{the specific details of how it can be implemented in the context of cognitive science models. The workflow is thus a general process that indicates how the user is responsible for making many choices (summary statistics, utility functions, etc.) appropriate for the context of their analysis. Our example analysis demonstrates just some possible choices.}

In order to fill this gap, for the field of cognitive science (linguistics, psychology, and related areas), we \textcolor{black}{illustrate one possible implementation of a principled Bayesian workflow with a reading-time experiment}.
\textcolor{black}{In a first part, we discuss the steps of the principled Bayesian workflow} (Betancourt, 2018). \textcolor{black}{This involves strategies for model building: we discuss an incremental model building strategy} (see Betancourt, 2018), \textcolor{black}{and we discuss a "maximal" model common in experimental contexts} (Barr, Levy, Scheepers, \& Tily, 2013). \textcolor{black}{Moreover, we describe the principled questions to be asked on the model} (cf. Betancourt, 2018), \textcolor{black}{and illustrate the underlying concepts visually} (cf. Betancourt, 2018).
\textcolor{black}{We then apply this principled Bayesian workflow to an example from the cognitive sciences (a psycholinguistic data set).} We demonstrate how to implement analyses in R, and use the R package \emph{brms} (Bürkner, 2017b) for statistical analysis. \textcolor{black}{Note that we illustrate these concepts using a linear (mixed) model, but that the same principles and workflow are also valid for non-linear cognitive process models.}

Note that some parts of this principled Bayesian workflow can demand considerable computational resources. Some of these checks, however, could be implemented once for a given research program, as similar experimental designs and statistical analyses may not demand a fundamental re-analysis of all the steps of this workflow for every single follow-up experiment.

Before starting with describing the workflow, we briefly lay out some basic definitions and terminology related to Bayesian modeling and inference. For a detailed introductory treatment, see Lambert (2018), and also \url{http://bit.ly/2GPDW74}.

In the \textcolor{black}{cognitive sciences, we use the} Bayesian framework \textcolor{black}{with the aim} to understand the processes that have generated some observed data \(y\). For this, we use a statistical (or possibly a computational) model \(\pi(y \mid \theta)\), \textcolor{black}{which quantifies a set of mathematical narratives for how the data} \(y\) \textcolor{black}{might be generated, each narrative labeled by a parameter value,} \(theta\). A simple example of a model could be a linear regression\textcolor{black}{, where the likelihood would be} \(\pi(y \mid \theta, \sigma) = \prod_i \frac{1}{\sigma \sqrt{2\pi}}e^{\frac{(\theta - y_i)^2}{-2\sigma^2}}\). Here, we consider the more complex case of a hierarchical linear model, familiar to cognitive scientists as the linear mixed model (Pinheiro \& Bates, 2000).

\textcolor{black}{When the observational model is evaluated at a particular data set it defines a likelihood function encoding which narratives are more consistent with the observed data than others.} \textcolor{black}{We write down the mathematical form of this more complex model below.} Bayes' rule allows us to use the likelihood to obtain the posterior probability distribution of the parameters given the data \(\pi(\theta \mid y)\):
\begin{equation}
\pi(\theta \mid y)=\frac{\pi (y \mid\theta) \pi(\theta)}{\pi(y)}
\end{equation}
This involves prior distributions over the parameters \(\pi(\theta)\). These prior distributions can represent pre-existing knowledge or beliefs about the parameters that are available before the data are observed. \textcolor{black}{Incorporating such principled domain expertise in the prior is of key importance, and provides a major advantage of Bayesian modeling.} \textcolor{black}{The term} \(\pi(y)\) is a normalizing constant, and is obtained \textcolor{black}{by integrating} over the whole parameter \textcolor{black}{space}: \(\pi(y) = \int \pi (y \mid \theta) \pi(\theta) \mathrm{d}\theta\). Because it is a constant, it can be ignored when computing the posterior distributions of the parameters.

Given the likelihood and the prior, a key challenge in Bayesian computations is to compute the posterior \textcolor{black}{distributions of the parameters accurately, and to compute posterior expectations from this}, for example the posterior mean, or alternatively posterior quantiles in credible intervals. For most interesting applications, the posterior expectations cannot be computed analytically. Instead, probabilistic posterior sampling (as implemented in various probabilistic programming languages such as Stan) is a method of choice for performing accurate posterior computations.

Here, we provide a detailed description of a number of questions to ask about a model, and checks to perform to validate a probabilistic model. Before going into the details of this discussion, we first treat the process of model building, and how different traditions have yielded different approaches to this questions.

\hypertarget{model-building}{%
\subsection{Model building}\label{model-building}}

One strategy for model building is to start with a minimal \textcolor{black}{observational model and complementary prior} model that captures just the phenomenon of interest but not much other structure in the data. \textcolor{black}{Note that the "model" is the combination of the observational model and the prior model.} For example, this could be a linear model with just the factor or covariate of main interest. For this model, we perform a number of checks described in detail in the following sections. If the model passes all checks and does not show signs of inadequacy, then it can be applied in practice and we can be confident that the model provides reasonably robust inferences on our scientific question. If the model shows signs of trouble on one or more of these checks, however, then the model may need to be improved. Alternatively, we may need to be more modest with respect to our scientific question. For example, in a repeated measures data-set, we may be interested in estimating the correlation parameter between two random effects (i.e., their random effects correlation) based on a sample of \(30\) subjects. If model analysis reveals that our sample size is not sufficiently large to estimate the random effects correlation reliably, then we may need to either increase our sample size, or give up our plan of analyzing the random effects correlation based on this data.

During the model building process, we make use of an aspirational model \(\pi_A\): we mentally imagine a model with all the possible details that the phenomenon and measurement process contain; i.e., we imagine a model that one would fit if there were no limitations in resources, time, mathematical and computational tools, subjects, and so forth. It would contain all systematic effects that might influence the measurement process. For example, influences of time or heterogeneity across individuals. This should be taken to guide and inform model development; such a procedure prevents random walks in model space during model development. Note that the model has to consider both the latent phenomenon of interest as well as the environment and experiment used to probe it.

The initial model \(\pi_1\), to the contrary, may only contain enough structure to incorporate the phenomenon of core scientific interest, but none of the additional aspects/structures relevant for the modeling or measurement.
\textcolor{black}{Manifestations of the additional, initially left-out structures, which reflect the difference between the initial} (\(\pi_1\)) \textcolor{black}{and the aspirational model} (\(\pi_A\)) \textcolor{black}{, can then be looked for in the workflow.}
These summary statistics can thus inform model expansion from the initial model \(\pi_1\) into the direction of the aspirational model \(\pi_A\). If the initial model proves \textcolor{black}{to be} inadequate, then the aspirational model and the associated summary statistics guide model development. If the expanded model is still not adequate, then another cycle of model development is conducted.

The range of prior and posterior predictive checks discussed in the following sections provide a principled Bayesian workflow of how this model expansion is done. The notion of \emph{expansion} is critical here. If an expanded model does not prove more adequate, one can always fall back to the previous model version. \textcolor{black}{Note that in the case of linear models, this expansion is similar to standard methods of forward selection. The notion of expansion, however, is more general, and generalizes also to non-linear and process-based models.}

As an alternative analysis strategy, a rich tradition in the cognitive and other experimental sciences relies on \enquote{maximal models} for a given experimental design (Barr et al., 2013). This maximal model contains all effects from experimental manipulations (main effects and interactions) as well as
\textcolor{black}{all within-subject and within-item variance components}
. That is, this model is maximal within the scope of a linear regression. Such a maximal model provides an alternative starting point for the principled Bayesian workflow. In this case, the focus does not lie so much on model expansion \textcolor{black}{unless we can identify strong deviations from the linear model assumptions}. Instead, core goals are to specify priors encoding domain expertise, and to ensure computational faithfulness, model sensitivity, and model adequacy.

Note that the maximal model \textcolor{black}{is} not maximal with respect to the actual data generating process. \textcolor{black}{Maximal linear} models are still bound by the linear regression structure and hence cannot capture effects such as selection bias in the data, dynamical changes in processes across time, or measurement error. At the same time this restricted class of (linear regression) models is exactly what packages like \emph{brms} \textcolor{black}{and \textit{rstanarm}} target, so it might also be \enquote{maximal} within the possibilities of such tools. Importantly, however, these \enquote{maximal} models are not the aspirational model, which is an image of the true data generating process. Indeed, aiming to formulate models closer to the aspirational model, which go beyond the linear modeling framework, may be one reason to consider investing in learning to express models directly within probabilistic programming languages such as Stan and JAGS instead of the limited range of \textcolor{black}{(linear and non-linear)} models provided in packages like \emph{brms} \textcolor{black}{and \textit{rstanarm}}.

Finally, we note that sometimes the results from the Bayesian workflow will show that our experimental design or data is not sufficient to answer our scientific question at hand. In this case, ambition need to be \textcolor{black}{tempered}, or new data needs to be collected, possibly with a different experimental design more sensitive to the phenomenon of interest. \textcolor{black}{Alternatively, researchers may try out a different analysis or formulate a different model with different assumptions. This may lead to investigation of a different scientific question than initially asked, or to a different answer to the same question.}

One important development in open science practices is preregistration of experimental analyses \textcolor{black}{and computational modeling approaches} (Lee et al., 2019) before the data are collected. This can be done using online platforms such as the \href{https://osf.io/}{Open Science Foundation} or \href{https://aspredicted.org/}{AsPredicted}. What information can or should one document in preregistration of the Bayesian workflow? If one plans on using the maximal model for analysis, then this maximal model, including contrast coding (Rabe, Vasishth, Hohenstein, Kliegl, \& Schad, 2020; Schad, Hohenstein, Vasishth, \& Kliegl, 2020), fixed effects, and random effects should be described. In the case of incremental model building, if a model isn't a good fit to the data, then any resulting inference will be limited if not useless (Lee et al., 2019), so a rigid preregistration is useless unless one knows exactly what the model is. Thus, the deeper issue with preregistration is that a model cannot be confirmed until the phenomenon \emph{and} experiment are all extremely well understood (Lee et al., 2019).
One practical possibility is to describe the initial and the aspirational model, and the incremental strategy used to probe the initial model to move more towards the aspirational model. Note that this can also include delineation of summary statistics that one plans to use for probing the tested models. Even if it is difficult to spell out the aspirational model fully, it can be useful to preregister the initial model, summary statistics, and the principles one intends to apply in model selection. \textcolor{black}{Moreover, many aspects of data preprocessing can be fixed; e.g., which region of interest to analyze, or which dependent variable to use.}
\textcolor{black}{Although} the maximal modeling approach clearly reflects confirmatory hypothesis testing, the incremental model building strategy towards the aspirational model may be seen as lying at the boundary between confirmatory and exploratory, and becomes more confirmatory the more clearly the aspirational model can be spelled out a priori.

\hypertarget{principled-questions-on-a-model}{%
\section{Principled questions on a model}\label{principled-questions-on-a-model}}

What characterizes a useful probabilistic model? First, a useful probabilistic model should be consistent with domain expertise. Second, it is key \textcolor{black}{that the model} allows accurate posterior approximation. Third, it must capture enough of the experimental design to give useful answers to our questions. Finally, a useful probabilistic model should be rich enough to capture the structure of the true data generating process needed to answer scientific questions.

So what can we do aiming to meet these properties of our probabilistic model? In the following, we will outline a number of analysis steps to take and questions to ask in order to improve these properties for our model (a more technical presentation is provided in Betancourt, 2018).

In a first step, we will use prior predictive checks to investigate whether our model is consistent with our domain expertise. Next, we will investigate computational faithfulness by studying whether posterior estimation is accurate. Third, we study model sensitivity and whether we can recover model parameters with the given design and model. As the last step in model validation, posterior predictive checks assess model adequacy for the given data-set, that is, they investigate whether the model captures the relevant structure of the true data generating process.

\hypertarget{prior-predictive-checks-checking-consistency-with-domain-expertise}{%
\subsection{Prior predictive checks: Checking consistency with domain expertise}\label{prior-predictive-checks-checking-consistency-with-domain-expertise}}

The first key question for checking the model is whether the model and the distributions of prior parameters are consistent with domain expertise. Prior distributions can be selected based on prior research or plausibility. For complex models, however, it is often difficult to know which prior distributions should be chosen, and what consequences distributions of prior model parameters have for expected data. A viable solution is to use prior distributions to simulate hypothetical data from the model and to check whether the simulated data are plausible and consistent with domain expertise, which is often much easier to judge compared to assessing prior distributions in complex models directly. This approach has been suggested in the \emph{device of imaginary results} by Good (1950).
In practice, this can be implemented \textcolor{black}{with} the following steps. Do the following \(N_{\mathrm{sim}}\) times:

\begin{enumerate}
\def\labelenumi{\arabic{enumi}.}
\tightlist
\item
  Using the prior \(\pi(\theta)\), randomly draw a parameter set \(\tilde{\theta}\) from it: \(\tilde{\theta} \sim \pi(\theta)\)
\item
  Use this parameter set \(\tilde{\theta}\) to simulate \(n\) hypothetical data points \(\tilde{y}\) from the model: \(\tilde{y} \sim \pi(y \mid \tilde{\theta})\).
\end{enumerate}

\noindent
This simulation method should result in a matrix that has dimensisons \(N_{\mathrm{sim}} \times n\).

\begin{figure}

{\centering \includegraphics{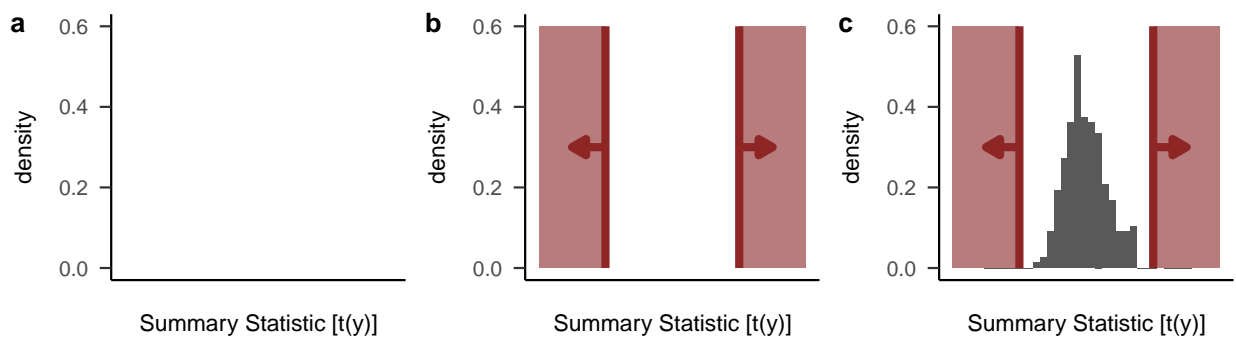} 

}

\caption{Prior predictive checks. a) In a first step, define a summary statistic that one wants to investigate. b) Second, define extremity thresholds (shaded areas), for which one does not expect a lot of prior data. c) Third, simulate prior model predictions for the data (histogram) and compare them with the extreme values (shaded areas).}\label{fig:FigPriorPredCh}
\end{figure}

To assess whether prior model predictions are consistent with domain expertise, it is useful to compute summary statistics of the simulated data \(t(\tilde{y})\). The distribution of these summary statistics can be visualized using, for example, histograms (see Fig.~\ref{fig:FigPriorPredCh}). This can quickly reveal whether the data falls in an expected range, or whether a substantial \textcolor{black}{number} of extreme data points are expected a priori. For example, in a study using self-paced reading times, \enquote{extreme} values may be considered to be reading times smaller than \(50\) ms or larger than \(2000\) ms, which would not be impossible, but would be implausible and largely inconsistent with domain expertise. A small number of observations may actually take extreme values. Observing a large number of extreme data points in the hypothetical data, however, would be inconsistent with domain expertise. In this case, the priors or the model should be adjusted to yield hypothetical data within a range of reasonable values.

\FloatBarrier

Choosing good summary statistics is more an art than a science. The choice of summary statistics will be crucial, however, as they provide key markers of what we want the model to account for in the data. They should thus be carefully chosen and designed based on expectations we have about the true data generating process and about the kinds of structures and effects we expect the data may exhibit. Interestingly, summary statistics can also be used to encode criticisms: if someone wants to criticize an analysis, then they can formalize that criticism into a summary statistic \textcolor{black}{that} they expect to show undesired behavior, which could e.g., provide a very constructive way to write reviews. Here, we will show some examples of useful summary statistics below when discussing data analysis for a concrete example data-set.

\begin{figure}

{\centering \includegraphics{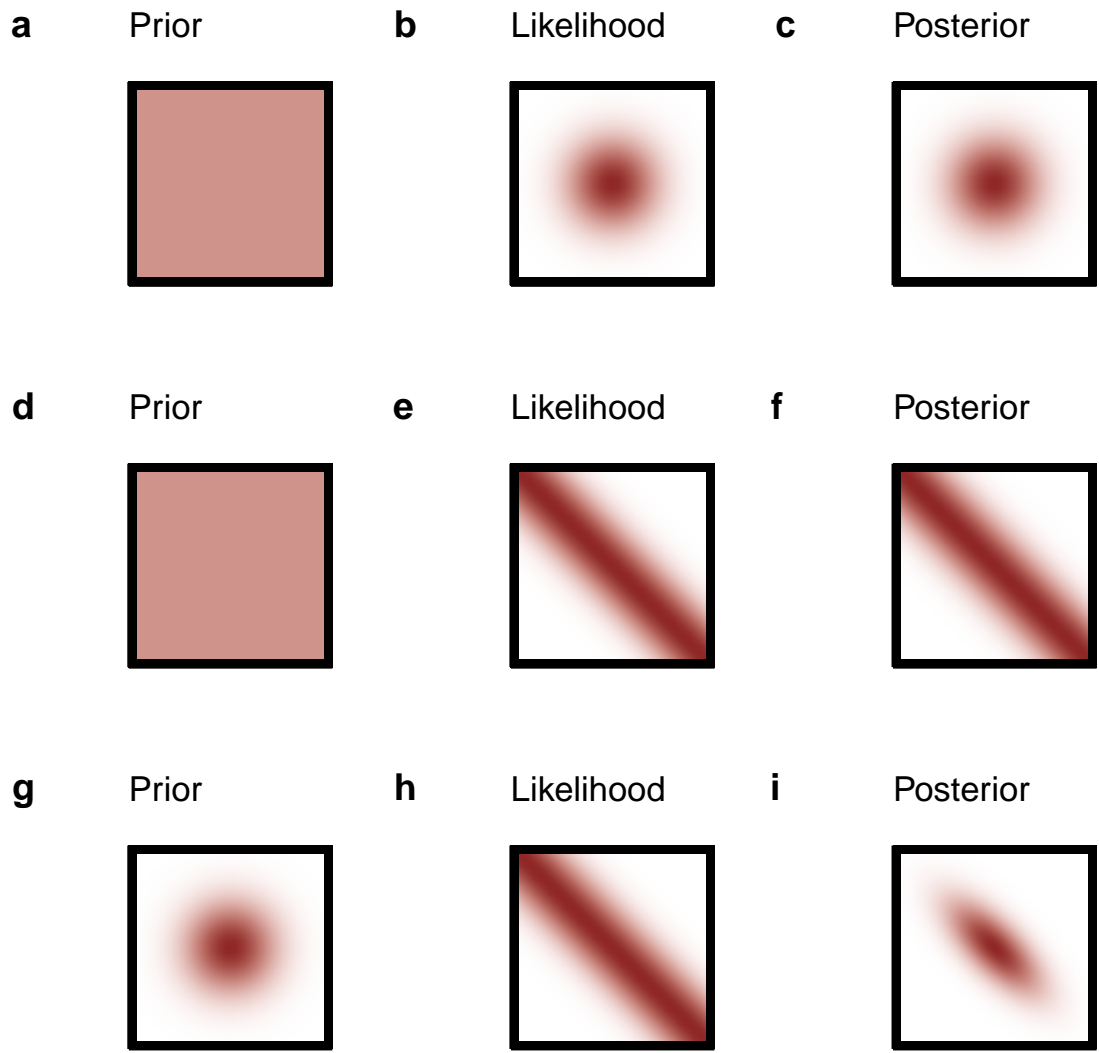} 

}

\caption{The role of priors for informative and uninformative data. a)-c) When the data provides good information via the likelihood (b), then a flat prior (a) is sufficient to obtain a concentrated posterior (c). d)-f) When the data does not sufficiently constrain the parameters through the likelihood (e), then using a flat prior (d) also leaves the posterior (f) diffuse. g)-i) When the data does not constrain the parameter through the likelihood (h), then including domain expertise into a (weakly) informative prior (g) can help to constrain the posterior (i) to reasonable values.}\label{fig:FigBayes}
\end{figure}

Choosing good priors will be particularly relevant in cases where the likelihood is not \textcolor{black}{"concentrated"} (see Fig.~\ref{fig:FigBayes}, in particular g-i). In linear mixed models, for example, this often occurs in cases where a maximal model is \textcolor{black}{fit to} a small data-set that does not constrain estimation of all \textcolor{black}{the} variance and covariance parameters \textcolor{black}{of the random effects}. In frequentist methods (such as implemented in the lme4 package in the lmer program), this \textcolor{black}{leads to convergence problems}, which indicate that the likelihood is too flat and that the parameter estimates \textcolor{black}{have high uncertainty}.

In such situations, using a prior in a Bayesian analysis (or a more informative prior rather than a diffuse one) should incorporate just enough domain expertise to suppress extreme \textcolor{black}{but} not impossible parameter values. This may allow the model to be fit, as the posterior is now sufficiently constrained.

\textcolor{black}{On} the contrary, frequentist linear model regression \textcolor{black}{theory} is built on assumptions about asymptotic properties of data sets. If the likelihood is not sufficiently informative to constrain the parameter values (such as in Fig.~\ref{fig:FigBayes}e), these asymptotic assumptions are invalid and the results of a frequentist linear model regression no longer fully characterize inferences about the model.
Therefore, introducing prior information in Bayesian computation allows fitting and interpreting models that cannot be validly estimated using frequentist tools.

A welcome side-effect of incorporating more domain expertise (into what still constitutes weakly informative priors) is thus more concentrated prior distributions, which can facilitate Bayesian computation. This allows more complex models to be estimated\textcolor{black}{;} that is, using prior knowledge can make it possible to fit models that could otherwise not be estimated using the available tools. In other words, incorporating prior knowledge allows us to get closer to the aspirational model in the iterative model building procedure. Moreover, more informative priors also lead to faster convergence of MCMC algorithms.

\FloatBarrier

Incorporating more domain expertise into the prior also has crucial consequences for Bayesian modeling when computing Bayes factors. Bayes factors are highly sensitive to the prior, and in particular to the prior uncertainty. Priors are thus never uninformative when it comes to Bayes factors. Choosing very diffuse priors (as in Fig.~\ref{fig:FigBayes}d) makes it very difficult to find posterior evidence in favor of an expanded model, and will often support the simpler model. For example, in nested model comparison of linear (mixed) models, large prior uncertainty (cf.~Fig.~\ref{fig:FigBayes}d) implies the assumption that the effect of interest could be very (implausibly) large. Using Bayes factors for such nested model comparison with high prior uncertainty thus tests whether there is evidence for a very big effect size for the predictor term in question, which is usually not supported by the data (because the diffuse prior covers implausibly large effect sizes). When using a weakly informative or even an informative prior (as in Fig.~\ref{fig:FigBayes}g), with much smaller uncertainty, to the contrary, the Bayes factor tests whether there is evidence for a small effect of the additional predictor term, which is much more likely to be the case. Thus, using prior knowledge and specifying priors with reasonable uncertainty (rather than diffuse priors with large uncertainty) are crucial in model comparison using Bayes factors.

The described process ends up simulating from the \emph{prior predictive distribution}, which specifies how the prior interacts with the likelihood. Mathematically, it computes an average (integral) over different possible (prior) parameter values. The prior predictive distribution is:
\begin{equation}
\pi(y) = \int \pi(y, \theta) \; \mathrm{d}\theta = \int \pi(y \mid \theta) \pi(\theta) \; \mathrm{d}\theta = \int \mathrm{likelihood}(y \mid \theta) \cdot \mathrm{prior}(\theta) \; \mathrm{d}\theta
\end{equation}
As a concrete example, suppose we assume that our likelihood is a Normal distribution with mean \(\mu\) and standard deviation \(\sigma\). Suppose that we now define the following priors on the parameters: \(\mu \sim Normal(0,1)\), and \(\sigma \sim Uniform(1,2)\). We can generate the prior predictive distribution using the following steps:

\begin{itemize}
\tightlist
\item
  Do the following 100,000 times:

  \begin{itemize}
  \tightlist
  \item
    Take one sample m from a Normal(0,1) distribution
  \item
    Take one sample s from a Uniform(1,2) distribution
  \item
    Generate and save a data point from Normal(m,s)
  \end{itemize}
\item
  The generated data is the prior predictive distribution.
\end{itemize}

\hypertarget{computational-faithfulness-testing-for-correct-posterior-approximations}{%
\subsection{Computational faithfulness: Testing for correct posterior approximations}\label{computational-faithfulness-testing-for-correct-posterior-approximations}}

A key aim in Bayesian data analysis is to compute posterior expectations, such as the posterior mean or posterior credible intervals (quantiles) of some parameter. For some simple models and prior distributions, these posterior expectations can be computed exactly by analytical derivation. This is not possible in more complex models, however, where analytical solutions cannot be computed. Instead, computational approximations are needed for estimation. One possible approximation is variational Bayes (MacKay, 2003), where parameterized probability density functions (e.g., the Gaussian distribution) are used for approximate posterior inference: the function parameters are estimated such that the function approximates the posterior as closely as possible. Here, we use another option: while it is often not possible to compute the posterior exactly, it is possible to draw samples from it, and we accordingly use (MCMC) sampling to approximate posterior expectations.

Approximations of posterior expectations can be inaccurate. For example, the computer program built to sample from a posterior can be erroneous. This could involve an error in the specification of the likelihood, or insufficient sampling of the full density of the posterior. The sampler may be biased by sampling parameter values that are larger or smaller than the true posterior, or the variance of the posterior samples may be larger or smaller than the true posterior uncertainty.

Given that posterior approximations can be inaccurate, it is important to design a procedure to test whether the posterior approximation of choice is indeed accurate, e.g., that the software used to implement the sampling works without errors for the specific problem at hand.

To test software, it is possible to use it in situations where there is a known \enquote{correct answer} and to compare the correct result with what the software generates. This approach, however, is more difficult for testing Bayesian estimation software, which can be stochastic. Here, an alternative can be to randomly sample model parameters from the prior distribution, then \textcolor{black}{simulate} data from the model (i.e., the likelihood function) given the sampled parameters, and perform Bayesian inference on the simulated data. If the Bayesian estimation software works correctly, then on average the obtained posterior will be correct. This means, for example, that over an ensemble of such simulations the true parameter values will be contained in \emph{any} 95\% posterior credible intervals in approximately 95\% of the simulations. (A 95\% credible interval indicates a Bayesian interval in which 95\% of the posterior probability mass is contained.) Note that there are many possible 95\% credible intervals --- within these simulations the average coverage will be 0.95 for all of them.

A powerful method for testing whether the posterior approximation of a software is correct is provided by simulation-based calibration (SBC) (Talts et al., 2018). It tests not only the correctness of 95\% or e.g., also of 75\% posterior credible intervals, but systematically tests the correctness of the whole posterior approximation. This is possible as it can be shown (see below) that \textcolor{black}{- and this is a remarkable and surprising fact -} on average the posterior looks like the prior. That is, if we repeatedly sample from the prior and then from the data, and compute many different posteriors from the \textcolor{black}{simulated} data, then the resulting ensemble of posteriors can be compared to the prior, and the average over the ensemble of posteriors should be the same as the prior. \textcolor{black}{This is a \textit{self consistency} condition and holds for any model. Because it holds for any model it doesn't provide any information about the validity of a particular model and hence can't be used to inform prior choice. All it does is allow us to check things in the internal context of a model.}

Mathematically, it can be shown that
\begin{equation}
\pi(\theta') = \int \int \pi(\theta' \mid y) \pi(y \mid \theta) \pi(\theta) \; \mathrm{d} y \mathrm{d} \theta
\end{equation}
\noindent
That is, we draw a sample \(\theta\) from the prior, \(\pi(\theta)\), simulate some data from the likelihood, \(\pi(y \mid \theta)\), and then estimate posterior parameters, \(\pi(\theta' \mid y)\), from the simulated data. When we take the average of the posterior over different simulated true parameters (\(\int \mathrm{d} \theta\)) and over different simulated data-sets (\(\int \mathrm{d} y\)), we will obtain a posterior that recovers the prior distribution (\(\pi(\theta')\)).

\textcolor{black}{SBC might seem similar to "parameter recovery". In Bayesian inference, however, there is no guarantee that the posterior will contain the true value for any single observation. "Parameter recovery" is a concern for calibration} \textcolor{black}{(see next section, on model sensitivity)} \textcolor{black}{where we verify that we have a model capable of learning enough from the data to recover parameters to the desired accuracy.
In contrast, SBC takes advantage of the fact that averaged over many observations the posterior distribution looks like the prior. We're not trying to recover parameters--we're testing the self-consistency of Bayesian inference.}

In practice, to conduct SBC, we use the following procedure:

\begin{enumerate}
\def\labelenumi{\arabic{enumi}.}
\tightlist
\item
  Take the prior \(\pi(\theta)\) and randomly draw a parameter set \(\tilde{\theta}\) from it: \(\tilde{\theta} \sim \pi(\theta)\)
\item
  Use this parameter set \(\tilde{\theta}\) to simulate hypothetical data \(\tilde{y}\) from the model: \(\tilde{y} \sim \pi(y \mid \tilde{\theta})\)
\item
  Fit the model to this hypothetical data and draw samples from the posterior distribution: \(\tilde{\theta}' \sim \pi(\theta \mid \tilde{y})\)
\end{enumerate}

We repeat steps 1 to 3 many times. In each cycle, we can compare posterior samples to the parameter set (from step 2.) used to simulate the hypothetical data. We record for each run where in the posterior distribution the prior parameters lie. If the distributions of the posterior samples and the sampled prior parameters are the same, the prior parameters should equally frequently lie at every location (i.e., rank) within the distribution of the posterior. Collecting all these locations gives an (ensemble) posterior sample of \(\tilde{\theta}'\) values.

\begin{figure}

{\centering \includegraphics{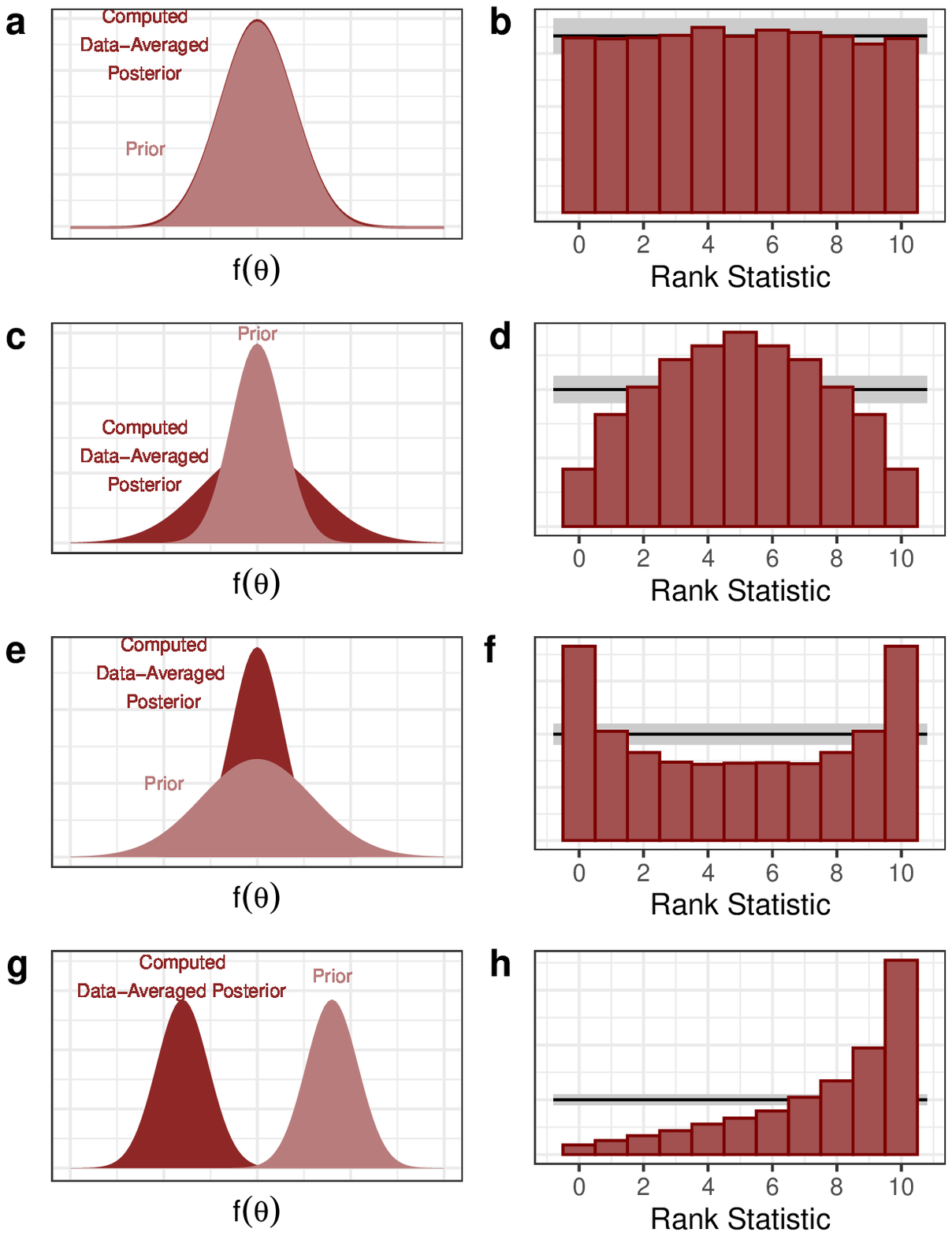} 

}

\caption{Exemplary results in simulation-based calibration (SBC). a)+b) If the data-averaged posterior exactly reflects the prior (identical prior and posterior; a) then the SBC histogram is uniformly distributed, indicating correct posterior approximation. b) The vertical line shows the expected number of ranks in each bin of the histogram based on a uniform distribution. c)-d) When the SBC histogram shows an inverse U-shaped form (d), then this indicates that the data-averaged posterior is over-dispersed (c), that is, that it has higher variance than the prior. e)-f) An SBC histogram showing a symmetric U-shape (f) indicates that the data-averaged posterior is under-dispersed (e), that is, it has lower variance than the prior. g)-h) If the SBC histogram is asymmetric (h), then the posterior will be biased in the opposite direction (g).}\label{fig:FigSBC4x2}
\end{figure}

\FloatBarrier

\begin{figure}

{\centering \includegraphics{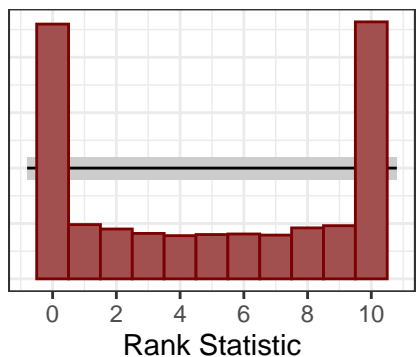} 

}

\caption{Exemplary results in simulation-based calibration (SBC): If the posterior samples exhibit considerable autocorrelation, then this can be visible as spikes at the boundaries of the histogram.}\label{fig:FigSBC1}
\end{figure}

\FloatBarrier

Accordingly, in simulation-based calibration (Talts et al., 2018), we take each simulated prior parameter value and test where it is located within the estimated posterior distribution by computing its rank statistic within the posterior. Said differently, we count the number of posterior parameter samples \(\tilde{\theta}_r'\) that are larger than the given prior parameter \(\tilde{\theta}\): \(\rho = \# \{ \tilde{\theta} < \tilde{\theta}_r' \}\).\footnote{\textcolor{black}{Note that an equivalent definition is possible by counting the number of posterior samples that are smaller (instead of larger) than the given prior parameter} (cf. Talts et al., 2018)\textcolor{black}{; this yields an equivalent result.}} We perform this calculation repeatedly, for every sampled prior parameter set. The resulting rank statistic of the prior parameters can be plotted as a histogram. As a central result of SBC (Cook, Gelman, \& Rubin, 2006; Talts et al., 2018), if the posterior model fitting works accurately, then the rank statistics are exactly uniformly distributed (see Fig.~\ref{fig:FigSBC4x2}a). Different patterns of how the distribution deviates from \textcolor{black}{the uniform distribution} can diagnose different specific problems of the posterior. We illustrate some examples (see Fig.~\ref{fig:FigSBC4x2} and Fig.~\ref{fig:FigSBC1}) (taken from Talts et al., 2018). Figure~\ref{fig:FigSBC4x2}c \textcolor{black}{shows a situation where the sampled posterior has a higher variance compared to the prior. This situation becomes evident as the histogram of SBC ranks takes an inverted U-shaped form. Alternatively, if the variance of the sampled posterior is too small (compared to the prior variance; cf.} Fig.~\ref{fig:FigSBC4x2}e), \textcolor{black}{then this is visible in a U-shaped form of the histogram of SBC ranks. If the sampled posterior distribution is biased compared to the prior, e.g., showing too small values }(see Fig.~\ref{fig:FigSBC4x2}g)\textcolor{black}{, then the SBC histogram of ranks will be asymmetric, with a lot of samples showing large ranks and only few samples showing small ranks. Last, if the posterior samples have high autocorrelation} (Fig.~\ref{fig:FigSBC1})\textcolor{black}{, then the SBC histogram of ranks will show a U-shaped form. Looking at the SBC histogram of ranks can thus provide some information about different ways of how the sampled posterior may deviate from the prior.}

Note that even powerful tools like Hamiltonian Monte Carlo (HMC) sampling do not always provide accurate posterior estimation, and that it's therefore important to check computational faithfulness for a given experimental design, model, and data-set. Only when we see that our posterior computations are accurate and faithful can we take the next step, namely looking at the sensitivity of the model analyses.

\hypertarget{model-sensitivity}{%
\subsection{Model sensitivity}\label{model-sensitivity}}

What can we realistically expect from the posterior of a model, and how can we check whether these expectations are justified for the current setup? First, we might expect that the posterior recovers the true parameters generating the data without bias. That is, when we simulate hypothetical data based on a true parameter value, we may expect that the posterior mean exhibits the same value. Although desirable, however, this expectation may or may not be justified for a given model, experimental design, and data-set. Indeed, parameter estimation for some, e.g., non-linear, models may be biased, such that the true value of the parameter can practically not be recovered from the data. At the same time, we might expect from the posterior that it is highly informative with respect to the parameters that generated the data. That is, we may hope for small posterior uncertainty relative to our prior knowledge, e.g., a small posterior standard deviation. Just as with the mean, however, the certainty in a posterior may not always be high. Some experimental designs, models, or data-sets may yield highly uninformative estimates, where uncertainty is not reduced compared to our prior information. \textcolor{black}{For example,} this can be the case when we have very little data, when the experimental design does not allow us to constrain certain model parameters\textcolor{black}{, or when we use a very complex model, where despite having a lot of data, the data may not be very informative for some of the complex model's parameters.}

\textcolor{black}{An example of the latter situation is the drift diffusion model} (Ratcliff \& Rouder, 2000)\textcolor{black}{, and the model parameter for the non-decision time. Even if very large amounts of data should be available for a number of subjects, this parameter essentially depends only on the shortest reaction times for each participant, and thus only on a very small number of data points. Thus, despite an overall very large amount of data, the posterior distribution for this parameter may still exhibit considerable posterior uncertainty.}

\textcolor{black}{Model sensitivity depends on the question being asked, which in general requires the specification of a utility function that quantifies how well a posterior distribution answers that question.  Because these questions, and hence utility functions, depend on the particular context of an application they must be customized for each application.}

\textcolor{black}{We can also complement these more precise questions, however, with some coarse questions that capture high-level information about the expected inferences and provide a useful general summary.}

\begin{enumerate}
\def\labelenumi{\arabic{enumi})}
\tightlist
\item
  How well does the estimated posterior mean match the true parameter \textcolor{black}{used for simulating the data}?
\item
  How much is uncertainty reduced from the prior to the posterior?
\end{enumerate}

First, to determine the distance of the posterior mean from the true simulating parameter\textcolor{black}{, scaled by the posterior variance}, it is possible to compute a \emph{posterior z-score}:
\begin{equation}
z = \frac{\mu_{post} - \tilde{\theta}}{\sigma_{post}} \label{eq:zScore}
\end{equation}
Here, the posterior mean \(\mu_{post}\) is compared to the true simulating parameter \(\tilde{\theta}\), and the difference between them is scaled by the posterior uncertainty (standard deviation, \(\sigma_{post}\)). This measure estimates how close the posterior mean is to the truth \textcolor{black}{relative to the posterior standard deviation, in other words how close the entire posterior distribution is to the truth value}. Small (absolute) values close to zero indicate that the posterior mean is close to the true parameter value, and large (absolute) values indicate that the posterior mean is far off the true generating model parameter. Note that large positive values versus large negative values indicate different positive versus negative biases in the posterior estimation.

Second, \emph{posterior contraction} estimates how much prior uncertainty is reduced in the posterior estimation:
\begin{equation}
s = 1 - \frac{\sigma_{post}^2}{\sigma_{prior}^2} = \frac{\sigma_{prior}^2 - \sigma_{posterior}^2}{\sigma_{prior}^2}
\end{equation}
Here, the variance of the posterior distribution, \(\sigma_{post}^2\), is divided by the prior variance, \(\sigma_{prior}^2\). In general, additional information from the likelihood will reduce uncertainty, such that the posterior variance will be smaller than the prior variance. If the data is highly informative, then the variance in the estimate is strongly reduced, and there will be strong posterior contraction \(s\) close to \(1\). When the data provide little information, however, then the posterior variance will be of similar size as the prior variance, and posterior contraction \(s\) will be close to \(0\).

\begin{figure}

{\centering \includegraphics{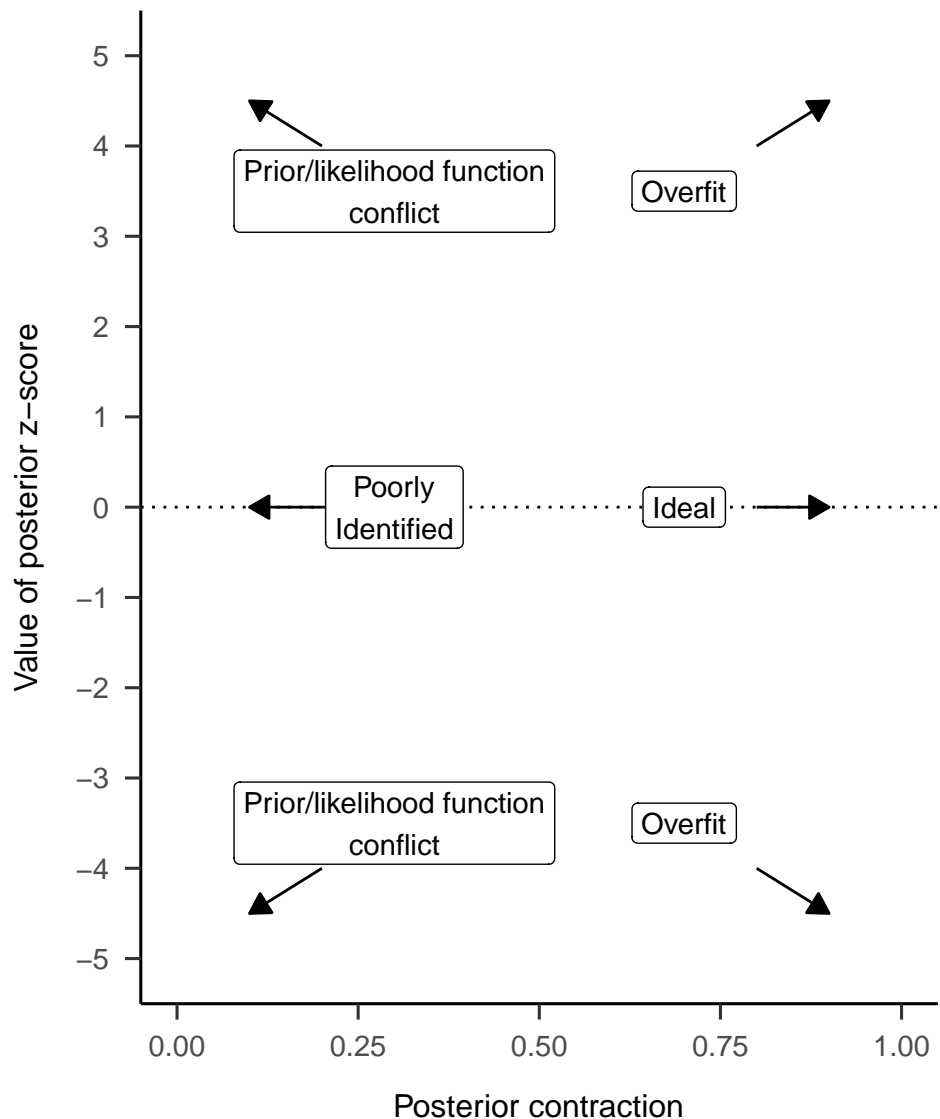} 

}

\caption{Posterior z-scores as a function of Posterior Contraction. Arrows show four possible results and their interpretation. The combination of high posterior contraction with large (positive or negative) posterior z-scores reflects situations of overfitting to noise in the data. Low posterior contraction with small z-scores reflect a poorly identified model. Low contraction with large (positive or negative) z-scores indicate a substantial conflict between the prior and the likelihood. Finally, high posterior contraction and low posterior z-scores reflect an ideal situation of good model fit.}\label{fig:FigSensitiv0}
\end{figure}

\FloatBarrier

To obtain these estimates, we take the following steps:

\begin{enumerate}
\def\labelenumi{\arabic{enumi}.}
\tightlist
\item
  Take the prior \(\pi(\theta)\) and randomly draw a parameter set \(\tilde{\theta}\) from it: \(\tilde{\theta} \sim \pi(\theta)\)
\item
  Use this parameter set \(\tilde{\theta}\) to simulate hypothetical data \(\tilde{y}\) from the model: \(\tilde{y} \sim \pi(y \mid \tilde{\theta})\)
\item
  Fit the model to this hypothetical data and draw samples from the posterior distribution: \(\tilde{\theta}' \sim \pi(\theta \mid \tilde{y})\)
\item
  Compute the \emph{posterior z-score} and the \emph{posterior contraction} for each sample of posterior parameters \(\tilde{\theta}'\).
\end{enumerate}

The distribution of \emph{posterior z-scores} and \emph{posterior contractions} can be plotted in a \textcolor{black}{two-dimensional grid as a} scatterplot (see Fig.~\ref{fig:FigSensitiv0}), which provides a useful model diagnostic: results in the upper \textcolor{black}{or lower} right corner \textcolor{black}{likely} indicate overfitting of the posterior to a wrong parameter value; results in the \textcolor{black}{middle} left indicate that the model is poorly identified, i.e., that the data do not well constrain estimation of model parameters; the upper \textcolor{black}{or lower} left corner reflects a situation of substantial conflict between the prior and the likelihood function; and the \textcolor{black}{middle} right \textcolor{black}{side} indicates the ideal situation of correct estimates with low uncertainty.

\textcolor{black}{Note that we choose rather large positive and negative z-values for the y-axis in Figure}~\ref{fig:FigSensitiv0}. \textcolor{black}{The reason of this is that small deviations of the estimated posterior mean are to be expected since the posterior is fitted onto simulated data, where the simulation process will introduce some noise and thus deviations of the estimated posterior mean. Larger z-scores, e.g., larger than absolute values of} \textcolor{black}{3 or 4,} however, \textcolor{black}{should only occur rarely due to this simulation process. Also, it is important to assess posterior z-scores for a range of simulated data sets. If no bias is present in the simulations, then the distribution of z-scores should be centered on $0$, whereas shifts in the distribution of z-scores to positive or negative values indicate a bias in the posterior estimation process.}

Importantly, the scatterplot doesn't provide a test for rejecting the current model. Instead its intent is to provide key information about the current setup, and to help us make a decision about the efficacy of the model, that is, whether the model allows us to obtain good estimates in this experimental design at all. Note that this can vary considerably over specific collected (or drawn) data-sets. Even with the model and the experimental design held constant, the model may be highly sensitive for some data-sets, but exhibit problems for another. This is one reason why we assess sensitivity for a range of different simulating parameter values covered by the prior distributions, to obtain information about the range of possible outcomes.

Note that this way of visualizing accuracy and contraction is just a general means of evaluating the utility of a model. More specific inferential goals can motivate more specific evaluations. For example, if we want to make a binary decision, \textcolor{black}{such as whether a parameter $\theta$ is larger than zero, $\theta > 0$}, then we might look at the distribution of false discovery rates and true discovery rates.

\hypertarget{posterior-predictive-checks-does-the-model-adequately-capture-the-data}{%
\subsection{Posterior predictive checks: Does the model adequately capture the data?}\label{posterior-predictive-checks-does-the-model-adequately-capture-the-data}}

\enquote{\emph{All models are wrong, but some are useful.}} (Box, 1979, p. 202) We know that our model probably does not fully capture the true data generating process, which is noisily reflected in the observed data. Our question therefore is whether our model is \emph{close enough} to the true process that has generated the data, and whether the model is useful for informing our scientific question. To compare the model to the true data generating process (i.e., to the data), we can simulate data from the model \textcolor{black}{fit to data} and compare the simulated to the real data. This can be \textcolor{black}{achieved} via a \emph{posterior predictive distribution}: the model is fit to the data, and the estimated posterior model parameters are used to simulate new data. The question then is how close the simulated data is to the observed data.

One way to assess this is using Bayesian cross validation or one of the many information criteria, such as BIC, DIC, or WAIC (e.g., Spiegelhalter, Best, Carlin, \& Linde, 2014; Shiffrin, Lee, Kim, \& Wagenmakers, 2008; Vehtari, Gelman, \& Gabry, 2017). This approach, however, only allows for relative comparison between different models, but not for an absolute measure of model fit. Moreover, the information criteria are only approximations and hence can be misleading when the approximation is inaccurate. The information criteria also consider the entire fit of the model, and hence can't differentiate between relevant aspects and irrelevant aspects.

\begin{figure}

{\centering \includegraphics{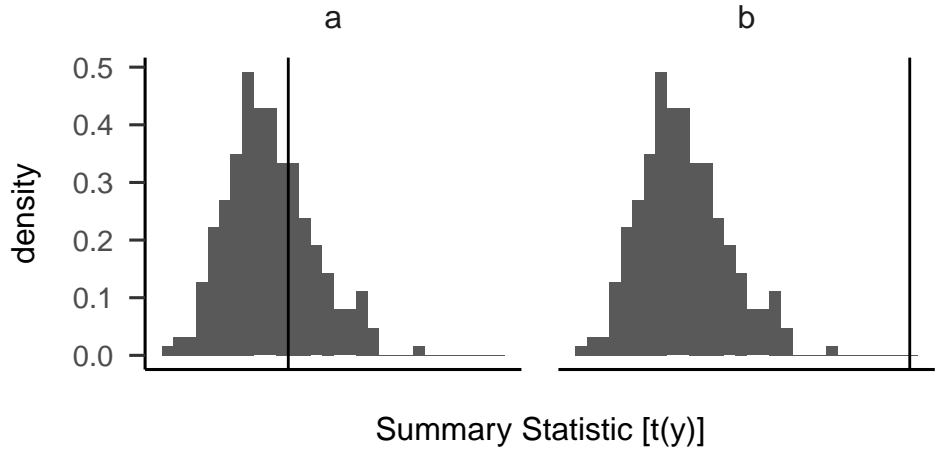} 

}

\caption{Posterior predictive checks. For a specific summary statistic, t(y), compare posterior model predictions (histogram) with the observed data (vertical line). a) This displays a case where the observed summary statistic (vertical line) lies within the posterior model predictions (histogram). b) This displays a case where the summary statistic of the observed data (vertical line) lies clearly outside of what the model predicts a posteriori (histogram).}\label{fig:FigPostPredCh}
\end{figure}

An alternative approach is to use features of the data that we care about, and to test how well the model can capture these. Indeed, we had already defined summary statistics in the prior predictive checks. We can now compute these summary statistics for the data simulated from the posterior predictive distribution. This will yield a distribution for each summary statistic. We compute the summary statistic for the observed data, and can now see whether the data falls within the distribution of the model predictions (cf.~Fig.~\ref{fig:FigPostPredCh}a), or whether the model predictions are far \textcolor{black}{from} the observed data (see Fig.~\ref{fig:FigPostPredCh}b). If the data is in the distribution of the model, then this supports model adequacy. \textcolor{black}{By contrast}, if we observe a large discrepancy, then this indicates that our model likely misses some important structure of the true process that has generated the data, and that we have to consider our domain expertise to further improve the model. Alternatively, however, a large discrepancy can be due to the data being an extreme observation, which was nevertheless generated by the process captured in our model. Note that in general we can't discriminate between these two possibilities. Consequently, we have to use our best judgement as to which possibility is more relevant, in particular changing the model only if the discrepancy is consistent with a known missing model feature.

\FloatBarrier

Mathematically, the \emph{posterior predictive distribution} is written:
\begin{equation}
\pi(y_{pred} \mid y) = \int \pi(y_{pred} \mid \theta) \pi(\theta \mid y) \; \mathrm{d} \theta 
\end{equation}
\noindent
Here, the observed data \(y\) is used to infer the posterior distribution over model parameters (\(\pi(\theta \mid y)\)). This is combined with the model or likelihood function (\(\pi(y_{pred} \mid \theta)\)) to yield new, simulated, data \(y_{pred}\). The integral \(\int \mathrm{d} \theta\) indicates averaging across different possible values for the posterior model parameters (\(\theta\)).

We can't evaluate this integral exactly: \(\theta\) can be a vector of many parameters, making this a very complicated integral with no analytical solution. However, we can approximate it using sampling. Specifically, we can obtain samples from the posterior distribution, e.g., using HMC or a different MCMC sampling scheme. We can now use each of the posterior samples as parameters to simulate new data from the model. This procedure then approximates the integral and yields an approximation to the posterior predictive distribution.

\hypertarget{exemplary-data-analysis}{%
\section{Exemplary data analysis}\label{exemplary-data-analysis}}

We perform an exemplary analysis of a data-set from Gibson \& Wu (2013). \textcolor{black}{We illustrate the principled Bayesian workflow for a linear mixed model. Note however, that the workflow is likewise valid for non-linear models of cognition. For the data by} Gibson \& Wu (2013), the methodology they used is called self-paced reading; this method is commonly used in psycholinguistics as a cheaper and faster substitute to eyetracking during reading. The participant is seated in front of a computer screen and is initially shown a series of broken lines that mask words from a complete sentence. The participant then unmasks the first word (or phrase) by pressing the space bar. Upon pressing the space bar again, the second word/phrase is unmasked and the first word/phrase is masked again; the time in milliseconds that elapsed between these two space-bar presses counts as the reading time for the first word/phrase. In this way, the reading time for each successive word/phrase in the sentence is recorded. Usually, the participant is also asked a yes/no question at the end of the trial, about the sentence. This is intended to ensure that the participant is adequately attending to the meaning of the sentence.

Gibson and Wu collected self-paced reading data using Chinese relative clauses. Relative clauses are sentences like: \emph{The student who praised the teacher was very happy}. Here, the \textcolor{black}{so-called} head noun, \emph{student}, is modified by a relative clause \emph{who\dots teacher}, and the head noun is the subject of the relative clause as well: the student praised the teacher. Such relative clauses are called subject relatives. By contrast, one can also have object relative clauses, where the head noun is modified by a relative clause which takes the head noun as an object. An example is: \emph{The student whom the teacher praised was very happy}. Here, the teacher praised the student.
Gibson and Wu were interested in testing the hypothesis that Chinese shows an object relative (OR) processing advantage relative to subject relatives (SR). The theoretical reasons for this are not interesting for the present purposes.
Their experimental design had one factor with two levels: (i) object relative sentence, and (ii) subject relative sentence. We use sum coding (\textcolor{black}{-1, +1}) for this factor, which we call \enquote{so}, an abbreviation for subject-object\textcolor{black}{, where we compute reading times in object relative sentences minus in subject relative sentences}. Following Gibson \& Wu (2013), we analyze reading time on a target word, which was the head noun of the relative clause; in Chinese, unlike English, the head noun appears after the relative clause. By the time the participant reads the head noun, they already know whether they are reading a subject or an object relative. The theory being tested here states that the meaning of the relative clause is resolved at the head noun and that in Chinese, this meaning resolution process is easier in object relatives vs.~subject relatives.

The data-set contains reading time measurements in milliseconds from \(37\) subjects and from \(15\) items. The design is a classic repeated measures Latin square design; there were originally \(16\) items, \textcolor{black}{but due to a scripting error, one item was not shown to participants. Three participants were removed because they had low question-response accuracy (less than 70\%; the remaining participants' accuracy was 91\%)}. We analyze the data using the R function \emph{brm} in the \emph{brms} package (Bürkner, 2017b), which provides an (\emph{lme4}-style syntax) interface to fit hierarchical (non-)linear models in the probabilistic programming language Stan (Carpenter et al., 2017).

\hypertarget{prior-predictive-checks}{%
\subsection{Prior predictive checks}\label{prior-predictive-checks}}

The first step in Bayesian data analysis is to specify the statistical model and the priors for the model parameters. In \emph{brms}, the latter can be done using the function \texttt{set\_prior()}. One possible standard setup for diffuse priors which is sometimes used in reading studies (e.g., Paape, Nicenboim, \& Vasishth, 2017; Vasishth et al., 2018a)\textcolor{black}{, but which we argue is an example of a "bad" prior}, is as follows: For the intercept we use a normal distribution with mean \(0\) and standard deviation \(10\). Note that this is on the log scale as we assume a lognormal distribution of reading times. That is, this approach assumes a priori that the intercept for reading times varies between \(0\) seconds and (one standard deviation) \(exp(10) = 22,026\) ms (i.e., \(22\) sec) or (two standard deviations) \(exp(20) = 485,165,195\) ms (i.e., \(135\) hours). Going from seconds to hours within one standard deviation shows how diffuse this prior is.
In \emph{brms} this is specified as: \texttt{set\_prior("normal(0,\ 10)",\ class\ =\ "Intercept")}. That is, we can specify a distribution (\texttt{normal(0,\ 10)}), and define the class of parameters for which this should apply; here, we specify \textcolor{black}{that} the prior should apply to the intercept (\texttt{class\ =\ "Intercept"}).

\begin{longtable}[]{@{}llll@{}}
\caption{\label{tab:EffectSize} Effect size as a function of the intercept.}\tabularnewline
\toprule
\begin{minipage}[b]{0.22\columnwidth}\raggedright
Intercept\strut
\end{minipage} & \begin{minipage}[b]{0.22\columnwidth}\raggedright
1 standard deviation below the mean\strut
\end{minipage} & \begin{minipage}[b]{0.22\columnwidth}\raggedright
1 standard deviation above the mean\strut
\end{minipage} & \begin{minipage}[b]{0.22\columnwidth}\raggedright
Effect size\strut
\end{minipage}\tabularnewline
\midrule
\endfirsthead
\toprule
\begin{minipage}[b]{0.22\columnwidth}\raggedright
Intercept\strut
\end{minipage} & \begin{minipage}[b]{0.22\columnwidth}\raggedright
1 standard deviation below the mean\strut
\end{minipage} & \begin{minipage}[b]{0.22\columnwidth}\raggedright
1 standard deviation above the mean\strut
\end{minipage} & \begin{minipage}[b]{0.22\columnwidth}\raggedright
Effect size\strut
\end{minipage}\tabularnewline
\midrule
\endhead
\begin{minipage}[t]{0.22\columnwidth}\raggedright
\(\exp(6) = 403\) ms\strut
\end{minipage} & \begin{minipage}[t]{0.22\columnwidth}\raggedright
\(\exp(6) / \exp(1) = 148\) ms\strut
\end{minipage} & \begin{minipage}[t]{0.22\columnwidth}\raggedright
\(\exp(6) \times \exp(1) = 1097\) ms\strut
\end{minipage} & \begin{minipage}[t]{0.22\columnwidth}\raggedright
\(1097 - 148 = 949\) ms\strut
\end{minipage}\tabularnewline
\begin{minipage}[t]{0.22\columnwidth}\raggedright
\(\exp(5) = 148\) ms\strut
\end{minipage} & \begin{minipage}[t]{0.22\columnwidth}\raggedright
\(\exp(5) / \exp(1) = 55\) ms\strut
\end{minipage} & \begin{minipage}[t]{0.22\columnwidth}\raggedright
\(\exp(5) \times \exp(1) = 403\) ms\strut
\end{minipage} & \begin{minipage}[t]{0.22\columnwidth}\raggedright
\(403 - 54 = 349\) ms\strut
\end{minipage}\tabularnewline
\begin{minipage}[t]{0.22\columnwidth}\raggedright
\(\exp(7) = 1097\) ms\strut
\end{minipage} & \begin{minipage}[t]{0.22\columnwidth}\raggedright
\(\exp(7) / \exp(1) = 403\) ms\strut
\end{minipage} & \begin{minipage}[t]{0.22\columnwidth}\raggedright
\(\exp(7) \times \exp(1) = 2981\) ms\strut
\end{minipage} & \begin{minipage}[t]{0.22\columnwidth}\raggedright
\(2981 - 403 = 2578\) ms\strut
\end{minipage}\tabularnewline
\bottomrule
\end{longtable}

For the effect of linguistic manipulations on reading times, one common standard prior is to assume a mean of \(0\) and a standard deviation of \(1\) (also on the log scale). Note that the prior on the effect size on log scale \textcolor{black}{specifies an effect size which} is a multiplicative factor, that is, the prediction for the effect size depends on the intercept.
For an intercept of \(exp(6) = 403\) ms, a variation to one standard deviation above multiples the base effect by \(2.71\), increasing the mean from \(403\) to \(exp(6) \times exp(1) = 1097\). Likewise a variation to one standard deviation below \textcolor{black}{divides the base effect by} \(2.71\), decreasing the mean from \(403\) to \(exp(6) \times exp(-1) = 148\) (see Table~\ref{tab:EffectSize}).
This effect size \textcolor{black}{changes dramatically} when assuming a different intercept: for a slightly smaller value for the intercept of \(exp(5) = 148\) ms, the expected condition difference is reduced to \(37\)\% (\(349\) ms), and for a slightly larger value for the intercept of \(exp(7) = 1097\) ms, the condition difference is enhanced to \(272\)\% (\(2578\) ms; see Table~\ref{tab:EffectSize}).
Here, we use this prior for the difference between object-relative and subject-relative sentences (i.e., the slope), and write this as \texttt{set\_prior("normal(0,\ 1)",\ class\ =\ "b",\ coef="so")}, where \texttt{class\ =\ "b"} indicates that all fixed effects share this prior, and \texttt{coef="so"} restricts it to the effect of the slope (OR - SR). \textcolor{black}{We use a truncated normal distribution with a mean of $0$ and a standard deviation of $1$ as the prior for the random effects standard deviations} (\texttt{class\ =\ "sd"}) \textcolor{black}{and for the residual variance} (\texttt{class\ =\ "sigma"}). \textcolor{black}{This can also be written in brms as} \texttt{normal(0,\ 1)}. \textcolor{black}{Because the prior is for a standard deviation, brm automatically uses a truncated normal distribution instead of the specified normal distribution} (\texttt{normal(0,\ 1)}). Finally, for the random effects correlation between the intercept and the slope, we use an \texttt{lkj} prior (Lewandowski, Kurowicka, \& Joe, 2009) with \textcolor{black}{a} diffuse prior parameter value of \(2\) (for visualization of the prior see Fig.~\ref{fig:FigLKJ}). We store these priors in an R object called \texttt{priors} (see Supplementary Code S1).

\begin{figure}

{\centering \includegraphics{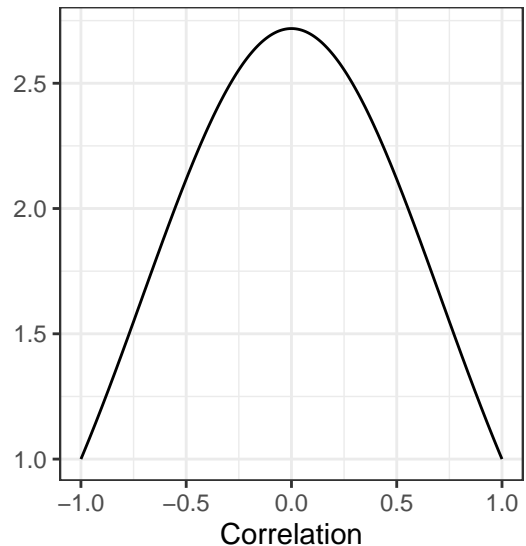} 

}

\caption{Shape of the LKJ prior with parameter 2. This is the prior density for the random effects correlation parameter, here used as a prior for the correlation between the effect size (so) and the intercept. The shape shows that correlation estimates close to zero are expected, and that very strong positive correlations (close to 1) or negative correlations (close to -1) are increasingly unlikely. Thus, correlation estimates are regularized towards zero.}\label{fig:FigLKJ}
\end{figure}

\FloatBarrier

For prior predictive checks, we use these priors to draw random parameter sets from the distributions, and to simulate hypothetical data using the statistical model. As a statistical model, we use the so-called maximal model (Barr et al., 2013) for the design. Such a model includes fixed effects for the intercept and the slope (\enquote{so}; coded using sum contrast coding: +1 for object relatives, and -1 for subject relatives), correlated random intercepts and slopes for participants (called subjects in the data frame), and correlated random intercepts and slopes for items. In \texttt{brms} syntax: \texttt{rt\ \textasciitilde{}\ 1+so\ +\ (1+so\textbar{}subj)\ +\ (1+so\textbar{}item)} and \texttt{family=lognormal()}.

\textcolor{black}{Mathematically, this model can be written as follows. We assume the reading time $RT_{sic}$ of subject $s$, item $i$, and experimental condition $c$ ("so") follows a log-normal distribution:}

\begin{equation}
p(RT_{sic} \mid \mu_{RT_{sic}}, \sigma_{RT}) = \frac{1}{RT_{sic}\cdot\sigma_{RT}\sqrt{2\pi}} \exp \{ -\frac{(ln RT_{sic}-\mu_{RT_{sic}})^2}{2\sigma_{RT}^2} \}
\end{equation}

\textcolor{black}{Here, $\mu_{RT,sic}$ is predicted for subject $s$, item $i$, and condition $c$. We assume that, $\mu_{RT,sic} = \beta_0 + S_{0,s} + I_{0,i} + (\beta_1 + S_{1,s} + I_{1,i}) \cdot so_{c}$. Here, $\beta_0$ is the intercept, $\beta_1$ is the slope (difference between subject and object relative sentences). The by-subject and by-item intercepts are $S_{0,s}$ and $I_{0,i}$, and the corresponding slopes are $S_{1,s}$ and $I_{1,i}$. The residual standard deviation is $\sigma_{RT}$.}

\textcolor{black}{The by-subject adjustments $S_s = \begin{pmatrix} S_{0,s} \\ S_{1,s} \end{pmatrix}$ follow a multivariate normal distribution with mean $\mu_0 = \begin{pmatrix} 0 \\ 0 \end{pmatrix}$ and variance-covariance matrix $\Sigma_S = \begin{pmatrix} \sigma_{S_0}^2 & \rho_S \sigma_{S_0} \sigma_{S_1} \\ \rho_S \sigma_{S_0} \sigma_{S_1} & \sigma_{S_1}^2 \end{pmatrix}$, where $\sigma_{S_0}$ is the random effects standard deviation of by-subject intercepts, $\sigma_{S_1}$ is the random effects standard deviation of by-subject slopes, and $\rho_S$ is the random-effects correlation between by-subject intercepts and slopes. 
The by-item adjustments $I_i = \begin{pmatrix} I_{0,i} \\ I_{1,i} \end{pmatrix}$ follow a multivariate normal distribution with mean $\mu_0 = \begin{pmatrix} 0 \\ 0 \end{pmatrix}$ and variance-covariance matrix $\Sigma_I = \begin{pmatrix} \sigma_{I_0}^2 & \rho_I \sigma_{I_0} \sigma_{I_1} \\ \rho_I \sigma_{I_0} \sigma_{I_1} & \sigma_{I_1}^2 \end{pmatrix}$, where $\sigma_{I_0}$ is the random effects standard deviation of by-item intercepts, $\sigma_{I_1}$ is the random effects standard deviation of by-item slopes, and $\rho_I$ is the random-effects correlation between by-item intercepts and slopes. The multivariate normal distributions are:}

\begin{align}
p(S_s \mid 0, \Sigma_S) &= (2\pi)^{-\frac{k}{2}}\det(\Sigma_S)^{-\frac{1}{2}} e^{-\frac{1}{2} S_s^T \Sigma_S^{-1} S_s } \\
p(I_i \mid 0, \Sigma_I) &= (2\pi)^{-\frac{l}{2}}\det(\Sigma_I)^{-\frac{1}{2}} e^{-\frac{1}{2} I_i^T \Sigma_I^{-1} I_i }
\end{align}

\textcolor{black}{We specify priors for all parameters in the model.}

\begin{itemize}

\item \textcolor{black}{For the intercept parameter $\beta_0$ we specify a normal distribution: $\beta_0 \sim N(\mu_0, \sigma_0)$. Here, we set $\mu_0$ to $0$, and we set $\sigma_0$ to $10$.}

\item \textcolor{black}{Likewise, for the slope parameter $\beta_1$ we specify a normal distribution: $\beta_1 \sim N(\mu_1, \sigma_1)$. We set $\mu_1$ to $0$, and we initially set $\sigma_1$ to $1$.}

\textcolor{black}{The other prior parameters specify the random effects terms in the model. These are discussed next.}

\item \textcolor{black}{For each of the standard deviations $\sigma_{re} = <\sigma_{S_0}, \sigma_{S_1}, \sigma_{I_0}, \sigma_{I_1}>$ of the random effect parameters varying across subjects ($\sigma_{S_0}$ and $\sigma_{S_1}$) and across items ($\sigma_{I_0}$ and $\sigma_{I_1}$), we specify a truncated normal $N_+$, or specifically a half-normal distribution with $\sigma_{re} > 0$: $\sigma_{re} \sim N_+(0, \sigma_{\sigma_{re}})$. Initially, we set $\sigma_{\sigma_{re}}$ to $1$ (and the mean of the truncated normal is assumed to be $0$).}

\item \textcolor{black}{For the random effects correlations $\rho = <\rho_S, \rho_I>$ between the intercept and the slope varying across subjects ($\rho_S$) and varying across items ($\rho_I$), we assume an LKJ prior distribution: $p(\rho \mid \eta) = Z(\eta) \cdot \det{|\rho|}^{\eta-1}$. We set the prior parameter $\eta$ to a value of $2$.}

\item \textcolor{black}{For the standard deviation of the log-normal distribution, $\sigma_{RT}$, we specify a truncated normal $N_+$, or specifically a half-normal distribution with $\sigma_{RT} > 0$: $\sigma_{RT} \sim N_+(0, \nu_{int})$. Initially, we set $\nu_{int}$ to a value of $1$ (and the mean of the truncated normal is assumed to be $0$).}

\end{itemize}

Next, we load the data to extract the experimental design. We use our custom R functions \texttt{SimFromPrior()} and \texttt{genfake()} to simulate parameters from the priors and to simulate data from the model (see Supplementary Code S2).

Based on the simulated data we can now perform prior predictive checks: we compute summary statistics, and plot the distributions of the summary statistic across simulated data-sets. First, we visualize the distribution of the simulated data. For a single data-set, this could be visualized as a histogram. Here, we have a large number of simulated data-sets, and thus a large number of histograms. We represent this uncertainty: for each bin, we plot the median as well as quantiles showing where 10\%-90\%, 20\%-80\%, 30\%-70\%, and 40\%-60\% of the histograms lie (see Supplementary Code S3).

\begin{figure}

{\centering \includegraphics{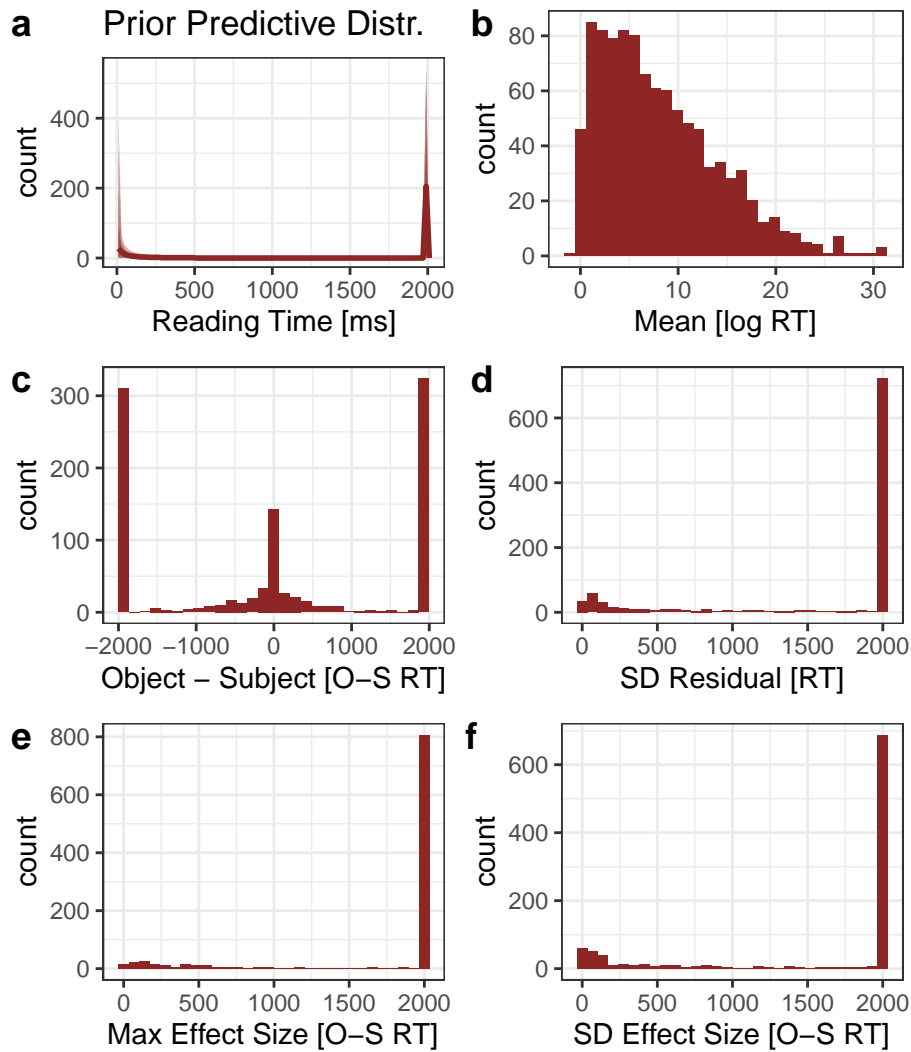} 

}

\caption{Prior predictive checks for a high-variance prior. Distributions are over simulated hypothetical data. a) Multivariate summary statistic: Distribution of histograms of reading times shows very short and also very long reading times are expected too frequently by the diffuse prior. b)-f) Scalar summary statistics. b) Distribution of average log reading times shows that extremely large reading times, e.g., of exp(10) = 22,026 ms are too frequently expected. c) Distribution of differences in reading times between object minus subject relatives shows that very large effect sizes are far too frequently expected. d) Distribution of standard deviations of residual reading times shows that very large variances are over-expected in the priors. e) Maximal effect size (object - subject relatives) across subjects again shows far too many extreme values. f) Standard deviation of effect size (object - subject relatives) across subjects; again far too many extreme values are expected. a)+c)-f) Values > 2000 or < -2000 are plotted at a value of 2000 or -2000 for visualization.}\label{fig:FigPrior1}
\end{figure}

For the current prior data simulations, this shows (see Figure~\ref{fig:FigPrior1}a) that most of the hypothetical reading times are close to zero or larger than \(2,000\) ms. It is immediately \textcolor{black}{clear} that the data predicted by this prior follows a very implausible distribution: it looks exponential, while we would expect a lognormal (or normal) distribution for reading times. Most data points take on extreme values. \textcolor{black}{Next, we look at further, univariate, summary statistics that characterize this distribution (e.g., mean and residual standard deviation, see below) and moreover look at other aspects of the model.}

As \textcolor{black}{two} additional summary statistics \textcolor{black}{of this distribution}, we take a look at the mean per simulated data-set (at the log scale) and also at the variance (ms scale; see Supplementary Code S4).
The results \textcolor{black}{for the mean}, displayed in Figure~\ref{fig:FigPrior1}b, show that the mean varies across a wide range, with a substantial number of data-sets having a mean larger than \(10\) on log scale or of \(exp(10) = 22,026\) on ms scale. Again, this reveals a highly unplausible assumption about the intercept parameter. The standard deviation, shown on ms scale for easier readability (Fig.~\ref{fig:FigPrior1}d), exhibits a substantial number of values larger than \(2,000\), which again is clearly larger than what we would expect for reading times.

We also plot the size of the effect of object relative minus subject relative sentence as a measure of effect size (Fig.~\ref{fig:FigPrior1}c; see Supplementary Code S5).
The results show that our priors commonly assume differences in reading times between conditions of more than \(2,000\) ms (see Fig.~\ref{fig:FigPrior1}c), which is larger than we would expect for a psycholinguistic manipulation of the kind investigated here. \textcolor{black}{Note that this is marginalizing the effect across different values for this intercept term.} More specifically, given that we model reading times using a lognormal distribution, the expected effect size depends on the value for the intercept. For example, for an intercept of \(exp(1) = 2.7\) ms and an effect size in log space of \(1\) (i.e., one standard deviation of the prior for the effect size), expected reading times for the two conditions are \(exp(1-1) = 1\) ms and \(exp(1+1) = 7\) ms. \textcolor{black}{By contrast,} for an intercept of \(exp(10) = 22,026\) ms, the corresponding reading times for the two conditions would be \(exp(10-1) = 8,103\) ms and \(exp(10+1) = 59,874\) ms.

Note that this implies highly varying expectations for the effect size, including the possibility for very large effect sizes. If we haven't seen an effect before running the experiment, however, then we would probably expect the effect to be rather small. Thus, priors with smaller expected effect sizes may be reasonable.

\textcolor{black}{Figure}~\ref{fig:FigPrior1}e \textcolor{black}{shows} individual differences in the effect of object versus subject relatives. \textcolor{black}{As a summary statistic (see Supplementary Code S6), we compute the effect size for} the participant with the largest (absolute) difference in reading times between object versus subject relatives (Fig.~\ref{fig:FigPrior1}e).
The prior simulations show maximal effect sizes of larger than \(2,000\) ms (Fig.~\ref{fig:FigPrior1}e), which is more than we would expect for observed data. Similarly, the variance in hypothetical effect sizes \textcolor{black}{across subjects} is large, with many SDs larger than \(2,000\) ms (Fig.~\ref{fig:FigPrior1}f), and thus again takes many values that are inconsistent with our domain expertise about reading experiments.

\FloatBarrier

\hypertarget{adjusting-priors}{%
\subsubsection{Adjusting priors}\label{adjusting-priors}}

Based on these \textcolor{black}{graphical summaries} of prior predictive data, we can use our domain expertise to refine our priors and adjust them to values for which we expect more plausible prior predictive hypothetical data as captured in the summary statistics.

\begin{figure}

{\centering \includegraphics{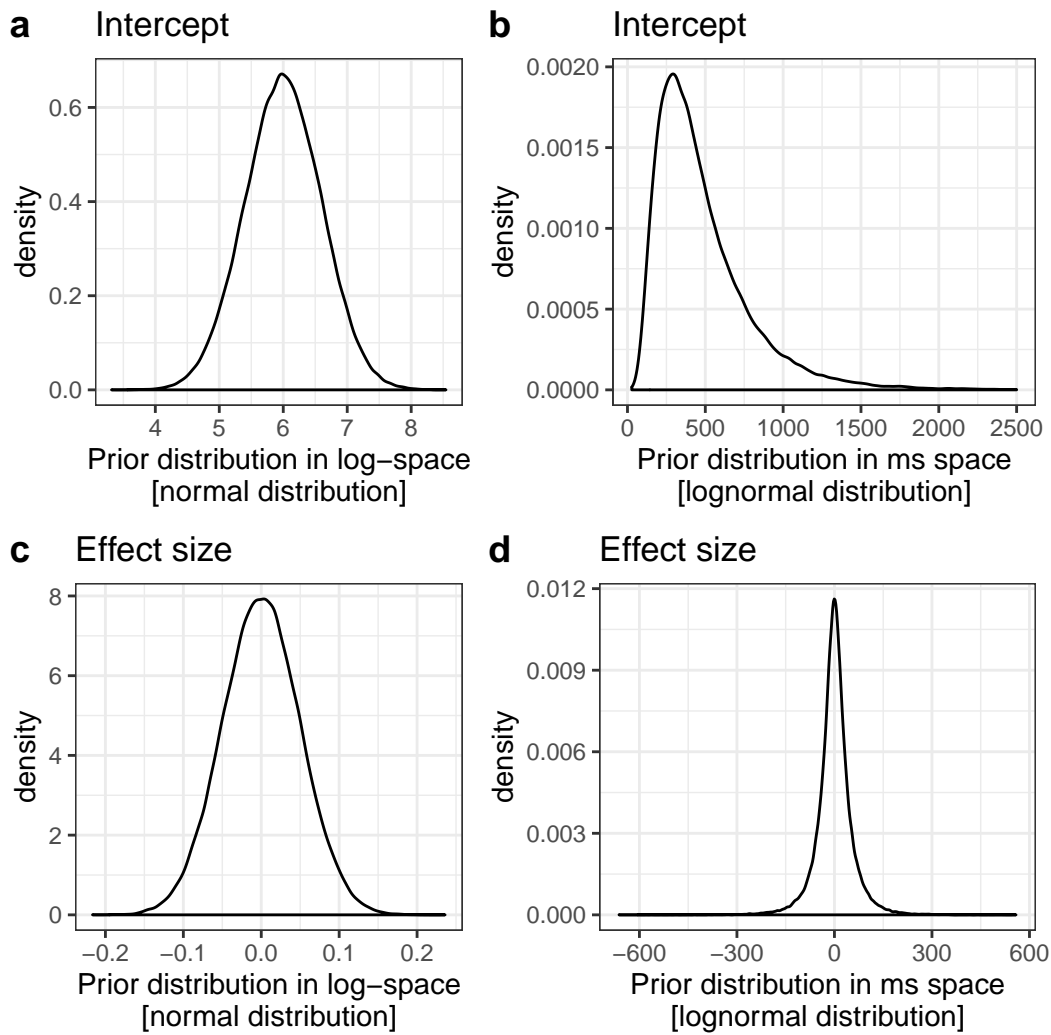} 

}

\caption{Prior distribution in log-space and in ms-space for a toy example of a linear regression. a) Displays the prior distribution of the intercept in log-space. b) Displays the prior distribution of the intercept in ms-space. c) Displays the prior distribution of the effect size in log-space, marginalizing over the intercept. d) Displays the prior distribution of the effect size in ms-space.}\label{fig:FigPriorAdjust1}
\end{figure}

First, we adapt the \textcolor{black}{prior on the} intercept. Upon re-consideration, we choose a normal distribution in log-space with a mean of \(6\). This corresponds to an expected grand average reading time of \(exp(6) = 403\) ms. For the standard deviation, we use a value of SD = \(0.6\). For these prior values, we expect a strongly reduced mean reading time and a strongly reduced standard deviation \textcolor{black}{of the residuals} in the simulated hypothetical data. Moreover, we expect that implausibly small or large values for reading times will no longer \textcolor{black}{occur}. For a visualization of the prior distribution of the intercept parameter in log-space and in ms-space, see Figure~\ref{fig:FigPriorAdjust1}a+b. \textcolor{black}{Slightly different} values, e.g., for the standard deviation \textcolor{black}{of the residuals} (e.g., SD = \(0.5\) or \(0.7\)), may yield similar results. Our goal is not to specify a precise value, but rather to use prior parameter values that are qualitatively in line with our domain expertise about expected observed reading time data, and that do not produce highly implausible hypothetical data.

\FloatBarrier

Next, for the effect of object minus subject relative sentences, we define a normally distributed prior with mean \(0\) and a much smaller standard deviation of \(0.05\). Again, we do not have precise information on the specific value for the standard deviation. We expect a generally smaller effect size (see the meta-analysis on Chinese relatives presented in Vasishth, Chen, Li, \& Guo, 2013), and we can check through prior predictive checks (data simulation and investigation of summary statistics) whether this yields a plausible pattern of expected results. Figures~\ref{fig:FigPriorAdjust1}c+d show expected effects in log-scale and in ms-scale for a simple linear regression example.

In addition, we assume much smaller values for the standard deviations in how the intercept and the slope vary across subjects and across items of \(0.1\), and a smaller standard deviation \textcolor{black}{of the residuals} of \(0.5\). For the correlation between random effects \textcolor{black}{we assume the same LKJ prior with the same parameter value of 2 as before}. For code summary, see Supplementary Code S7.

\hypertarget{prior-predictive-checks-for-weakly-informative-priors}{%
\subsubsection{Prior predictive checks for weakly informative priors}\label{prior-predictive-checks-for-weakly-informative-priors}}

Based on this new set of now weakly informative priors, we can again perform prior predictive checks. We again randomly draw samples of parameters from the priors, use these to simulate data from the statistical model, and compute summary statistics for the simulated data. We do not show the R code again for these analyses.

\begin{figure}

{\centering \includegraphics{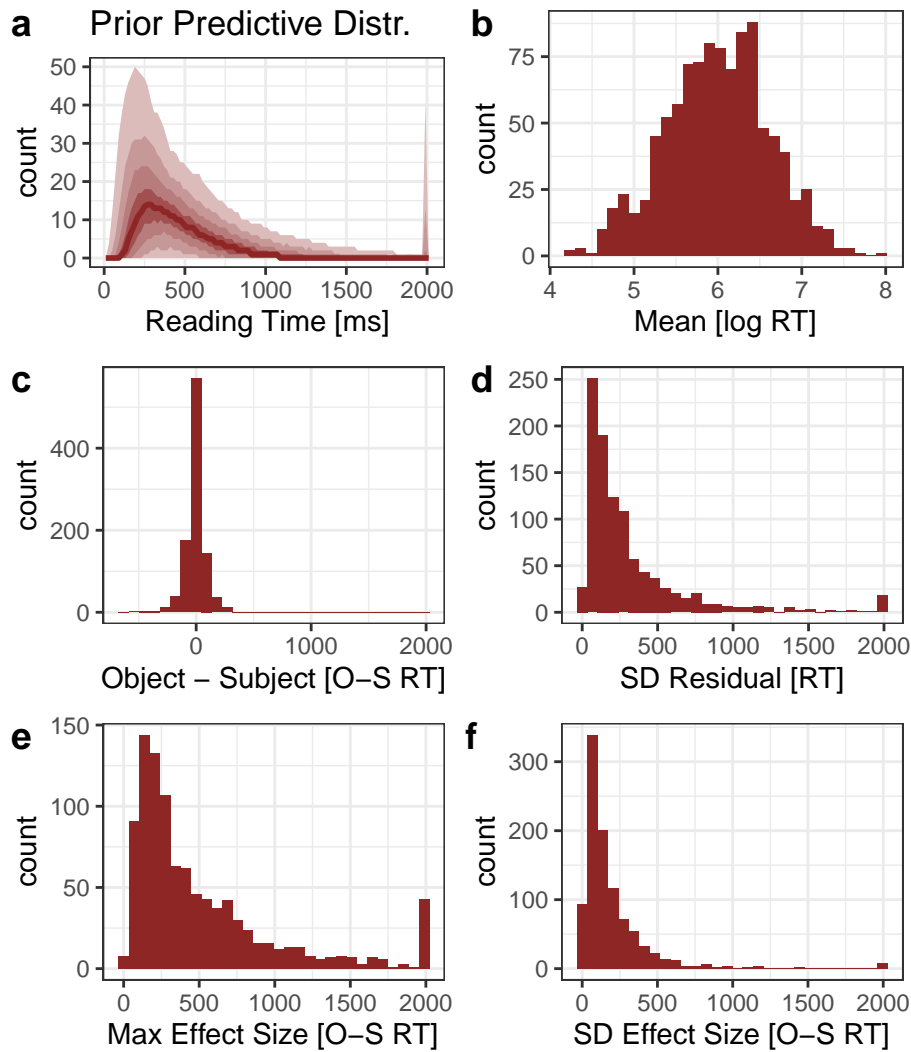} 

}

\caption{Prior predictive checks for weakly informative priors. Distributions are over simulated hypothetical data. a) Histograms of reading times. Shaded areas correspond to 10-90 percent, 20-80 percent, 30-70 percent, and 40-60 percent quantiles across histograms; the solid line (in the middle of the shaded area) indicates the median across hypothetical data-sets. This now provides a much more reasonable range of expectations. b) Average log reading times now span a more reasonable range of values. c) Differences in reading times between object minus subject relatives; the values are now much more constrained without many extreme values. d) Standard deviations of residual reading times; in contrast to the diffuse priors, values are in a reasonable range. e) Maximal effect size (object - subject relatives) across subjects; again, prior expectations are now much more reasonable compared to the diffuse prior. f) Standard deviation of effect size (object - subject relatives) across subjects; this no longer shows a dominance of extreme values. a)+c)-f) Values > 2000 or < -2000 are plotted at 2000 or -2000 for visualization.}\label{fig:FigPrior2}
\end{figure}

\FloatBarrier

Figure~\ref{fig:FigPrior2}a shows that now the distribution over histograms of the data looks much more reasonable, i.e., more like what we would expect for a histogram of observed data. Very small values for reading times are now rare, and not heavily inflated any more. Moreover, extremely large values for reading times larger than \(2,000\) ms are rather unlikely.

We also take a look at the hypothetical average reading times (in log space; Fig.~\ref{fig:FigPrior2}b), and find that our expectations are now much more reasonable. We expect average reading times of around \(log(6) = 403\) ms. Most of the expected average reading times lie between \(log(5) = 148\) ms and \(log(7) = 1097\) ms, and few extreme values beyond these numbers are observed. The standard deviations of the \textcolor{black}{residual} reading times are also in a much more reasonable range (see Fig.~\ref{fig:FigPrior2}d), with only very few values larger than the extreme value of \(2,000\) ms.

As a next step, we look at the expected effect size (OR minus SR) in the hypothetical data (Fig.~\ref{fig:FigPrior2}c). \textcolor{black}{In this analysis, we again marginalize over the intercept, that is, we consider all possible different values of the intercept in the computation}. Extreme values of larger or smaller than \(2,000\) ms are now very rare, and most of the absolute values of expected effect sizes are smaller than \(200\) ms. More specifically, we also check the maximal effect size among all subjects (\textcolor{black}{by computing the difference between mean reading times in subject versus object relative sentences for each subject, and taking the maximal absolute value}; Fig.~\ref{fig:FigPrior2}e). Most of the distribution centers below a value of \(1000\) ms, reflecting a more plausible range of expected values. Likewise, the standard deviation of the psycholinguistically interesting effect size \textcolor{black}{across subjects} now rarely takes values larger than \(500\) ms (Fig.~\ref{fig:FigPrior2}f), reflecting more modest assumptions than our first diffuse prior.

\hypertarget{computational-faithfulness}{%
\subsection{Computational faithfulness}\label{computational-faithfulness}}

\textcolor{black}{Next, w}e investigate the computational faithfulness of our computational methods for estimating posterior model parameters for the current experimental design and priors. In Stan and \emph{brm}, lack of divergences and Rhats close to 1 indicate no problem for each individual posterior fit \textcolor{black}{(the recommended cut-off is 1.05)}.

\begin{figure}

{\centering \includegraphics{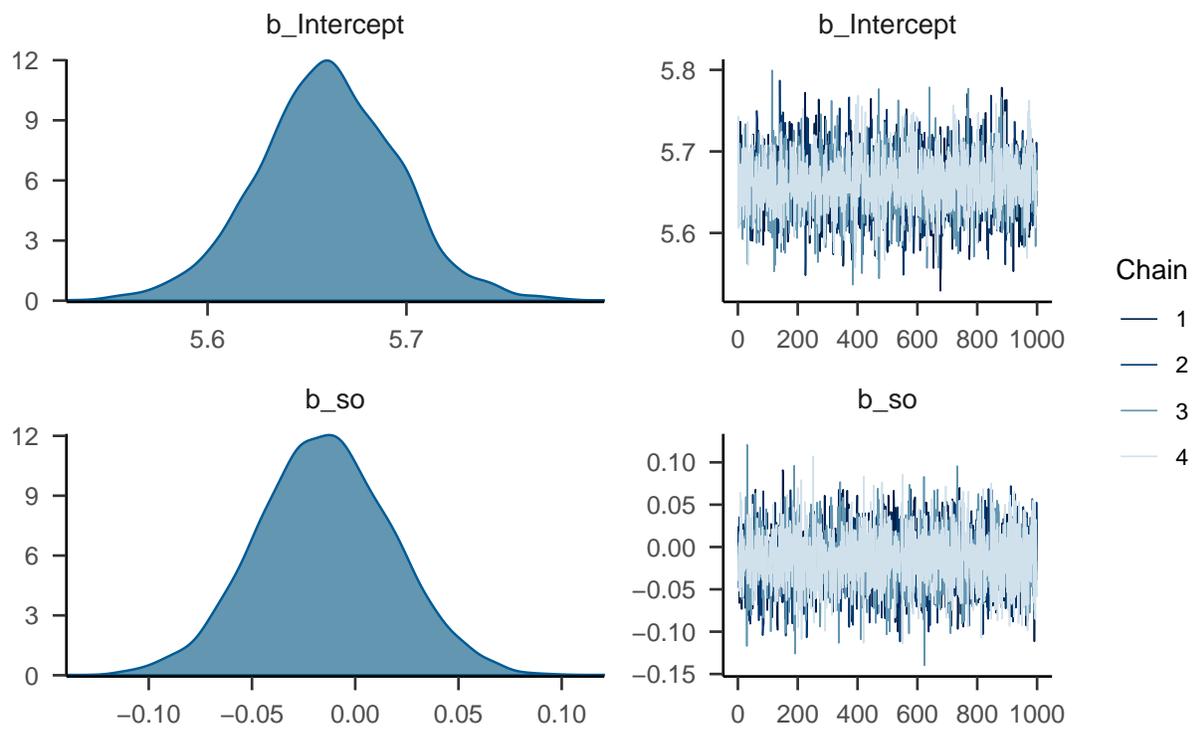} 

}

\caption{Density plots (left panels) and trace plots (right panels) for the intercept parameter (upper panels) and for the difference between subject and object relative sentences, labeled b so (lower panels).}\label{fig:TracePlots}
\end{figure}

\textcolor{black}{Rhat, however, is known to be a limited measure of convergence as it fails to detect some types of convergence problems} (Vehtari, Gelman, Simpson, Carpenter, \& Bürkner, 2019)\textcolor{black}{. Thus, even with a Rhat close to 1, one cannot be sure that the chains have converged. One alternative method is to inspect the trace plots visually (see Fig.}~\ref{fig:TracePlots}\textcolor{black}{), and to look for aspects of the samples that look like there could be problems with the sampling. Problems can be visible as high correlations between parameters, that is, two traces moving together over time. Alternatively, parameters may exhibit shifts over time, indicating their estimate has not yet stabilized, parameter values may be higher or lower for one chain than another, or parameters may not jump back and forth, but get stuck at certain values. All of these indicate problems with convergence.} \textcolor{black}{Trace plots, however, are impractical for the large models we want to build.} \textcolor{black}{Moreover, an improved version of Rhat is available together with more advanced plotting facilities to detect convergence issues} (Vehtari et al., 2019). \textcolor{black}{One difficulty with these kinds of criteria is that they generally cannot guarantee convergence. They can indicate certain problems with convergence. If no issues are found, however, this doesn't guarantee that other problems might still be present.}

\textcolor{black}{By contrast, SBC provides a way to test whether the posterior is computed accurately.} It aggregates the results from many simulations together to see if the ensemble shows any indication of inaccurate computation.
To investigate accuracy of computations, we use the simulated data drawn from the prior distributions and fit the statistical model to this simulated data, estimating (approximate) posterior distributions. We use the function \emph{brm} from the \texttt{brms} package for model fitting. Note that this process can take a considerable amount of time and computational resources. \textcolor{black}{On a single desktop machine with reasonably complex data, this can take days or even weeks.} Parallelizing model fits to multiple cores on a computing cluster can help bring computing time down. \textcolor{black}{If simulations take too long, the researcher may consider using a smaller number of simulations.} Talts et al. (2018) \textcolor{black}{suggest that any simulations are better than no simulations; thus, their recommendation is to do as many simulations as possible with the resources available.} \textcolor{black}{Because it is not practical to perform SBC} for every single model and every single analysis, an alternative path can be to investigate computational faithfulness (and model sensitivity) once for a given research program, where many aspects of the experimental design, the statistical model, and the priors may be repeated across analyses. For example, here we perform the analysis for the Gibson \& Wu (2013) design, and could also use these analyses when analyzing a replication of their experiment. First, we estimate the model for all \(1,000\) simulated data-sets (see Supplementary Code S8).

Next, we extract the number of divergent transitions of the HMC sampler to diagnose potential problems in model fitting (see Supplementary Code S9).

\begin{figure}

{\centering \includegraphics{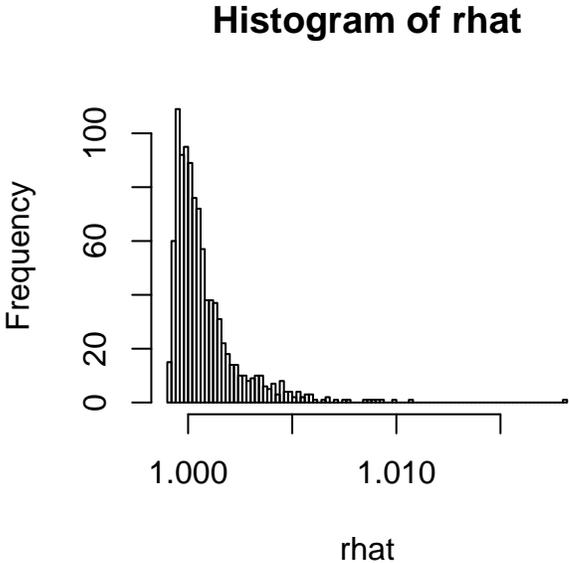} 

}

\caption{All values of rhat are close to 1, which indicates good model convergence for all of the fitted models.}\label{fig:FigRhat}
\end{figure}

We see that none of the models exhibited a difficulty with divergent samples. Divergent transitions indicate problems in the model fitting, and should be diagnosed and removed. In the present case, setting a control parameter known as \enquote{adapt\_delta} to a value higher than its default of 0.80 in \texttt{brms} removed the divergent transitions.
Another convergence diagnostic, \(\hat{R}\), is very close to \(1\) in all of the models (see Fig.~\ref{fig:FigRhat}), indicating no problems in model convergence.

\FloatBarrier

\FloatBarrier

The above steps \textcolor{black}{provide an indication} that the \textcolor{black}{HMC chains may have converged to the true posterior distribution and may have mixed well. This is visible in the right panels of Figure}~\ref{fig:TracePlots}\textcolor{black}{. If the chains have converged and mixed well, then this plot looks like a "fat hairy caterpillar".}

Next, we perform simulation-based calibration (SBC) (Talts et al., 2018). This method aims to determine whether estimated posterior model parameters follow the same distribution as the prior model parameters used to generate the data. It does so by comparing posterior estimates with the prior parameters used for simulating data. We performed simulation-based calibration on the current data-set by computing, for each simulated data-set, the rank of the prior parameter sample within the posterior samples. More specifically, we compute the number of posterior samples that are larger than the prior (simulating) parameter \textcolor{black}{(i.e., larger than the sampled parameter value that was used to simulate the data on which the model was fit)}. If the posterior distributions accurately estimate the distribution of the parameter, then their distribution should be the same as the distribution of the actual parameters used to generate the data, and the ranks should be uniformly distributed.

Note that SBC presumes that posterior samples are independent and not correlated (see Fig.~\ref{fig:FigSBC1}). Our HMC samples, however, do exhibit autocorrelation. \textcolor{black}{Autocorrelation is the correlation of the samples with a delayed copy of itself, which can be computed for different delays} (Kmenta, 1971)\textcolor{black}{. Autocorrelation is visible in SBC histograms (see Fig.}~\ref{fig:FigSBC1}). To remove the autocorrelation from the posterior samples, we thin our samples by taking only every eighth sample (for a discussion of thinning in SBC see Talts et al., 2018). To test how many samples should be removed, it's possible to investigate the autocorrelation in the samples (see Supplementary Code S10).

\textcolor{black}{Importantly, note that thinning is usually not necessary nor advantageous when evaluating the final model for the data. It can be shown that posterior inference is more precise when using un-thinned samples as compared to thinned samples} (Link \& Eaton, 2012)\textcolor{black}{, and thus thinning reduces the precision of the analysis.}

Next, we plot a histogram of SBC ranks (compare Fig.~\ref{fig:FigSBC4x2}; see Supplementary Code S11).

\begin{figure}

{\centering \includegraphics{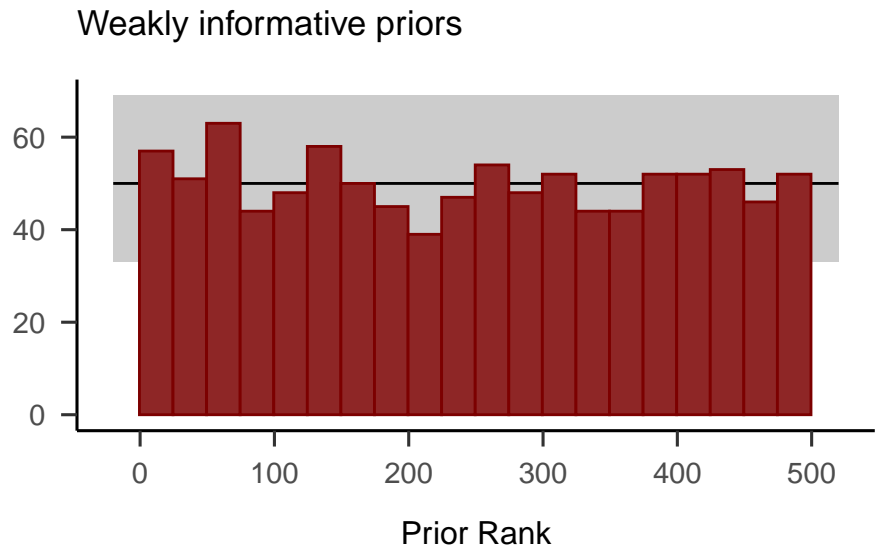} 

}

\caption{Simulation-based calibration. Histogram of simulation-based ranks of true parameters used for simulating data (randomly drawn from the prior distributions) within posterior samples fit to the simulated data. Shown for weakly informative priors. The results show that the red bars are all contained within the grey horizontal bar in the background, indicating the SBC ranks are uniformly distributed. This shows that the posterior recovers the prior distribution well and that the posterior is thus estimated accurately.}\label{fig:FigSBC}
\end{figure}

\FloatBarrier

In SBC, good recovery of the true simulating parameters in a posterior analysis is evident when the SBC ranks are uniformly distributed. Figure~\ref{fig:FigSBC} shows the histogram of SBC ranks \textcolor{black}{for the fixed effects parameter capturing the difference between subject and object relative sentences ("so")}. The grey bar in the background reflects the range of values to be expected based on a uniform distribution \textcolor{black}{(computed based on the quantile function of a binomial distribution)}. The results do not show any evidence of divergence from the uniform distribution. This result suggests that we can trust the posterior estimates from the \emph{brm} function for the current experimental design, statistical model, and prior distributions, that posterior samples do not exhibit bias, but instead that the samples from the posterior follow the same distribution as the prior.

\begin{figure}

{\centering \includegraphics{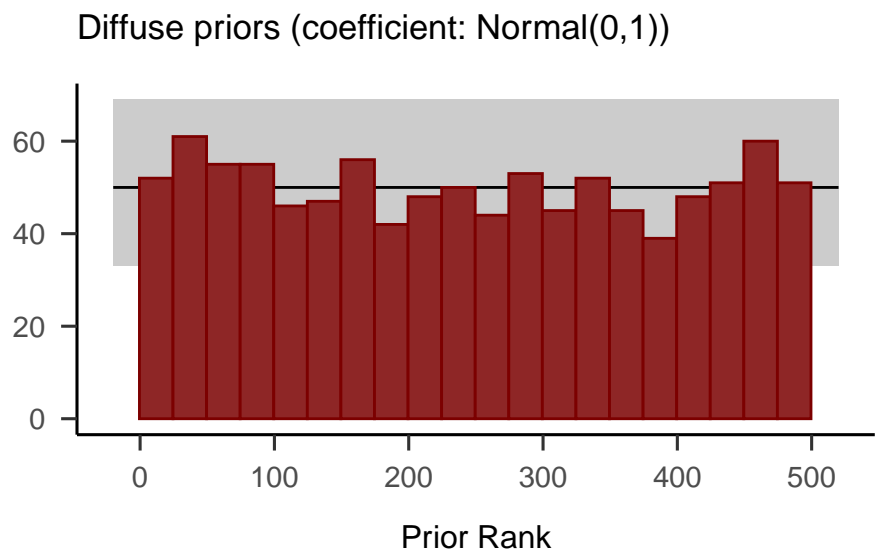} 

}

\caption{Simulation-based calibration for diffuse priors (intercept: Normal(0,10); coefficients: Normal(0,1)). This plot shows that, as for weakly informative priors, the SBC samples are uniformly distributed. This demonstrates that the computational methods work accurately also for the more diffuse priors. We saw above, however, that diffuse priors make very implausible assumptions.}\label{fig:FigSBCb}
\end{figure}

Note, however, that good recovery of model parameters is also possible when using more diffuse priors. As discussed above, diffuse priors have been used by many previous reading studies, choosing e.g., as the prior for the size of an experimental effect (here SR versus OR) a normal distribution with mean zero and standard deviation one. Figure~\ref{fig:FigSBCb} shows that such diffuse priors support similarly accurate posterior computations.

\FloatBarrier

Note that here we only investigated the slope parameter that estimates the difference in reading times between object relative and subject relative sentences (in log space). This is the parameter of theoretical interest to us. Similar analyses, however, are possible for all other model parameters to investigate whether the given computational methods provide accurate posterior estimates. Examples \textcolor{black}{of} other parameters of interest could be the standard deviation of the experimental effect across subjects, or the correlations of the effect with the intercept across subjects or items.
If the researcher actually cares only about one phenomenon, then the accuracy of that one effect is all one needs to check. That is, as long as the model provides good faithfulness for the effect of interest, it does not matter or hurt too much if other and irrelevant effects in the model are estimated more poorly (Betancourt, 2018).

Last, computational faithfulness seems to be an issue in frequentist approaches to standard linear mixed models (LMMs), where maximal LMMs frequently encounter difficulties with model fitting. \textcolor{black}{By contrast}, the HMC methods implemented in Stan and accessed using the brm function may be well able to cope with such models (even when using rather vague priors), such that computational faithfulness may be less of an issue with standard LMMs here. In brm the formulation of the likelihood is moreover highly standardized, preventing errors in its formulation. Testing computational faithfulness, moreover, will become very important when we define more customized models. \textcolor{black}{The utility to test computational faithfulness} is exactly one of the advantages of this framework (Nicenboim \& Vasishth, 2018).

\hypertarget{model-sensitivity-1}{%
\subsection{Model sensitivity}\label{model-sensitivity-1}}

\textcolor{black}{So far, we have identified appropriate priors using prior predictive checks, and have validated posterior estimates using SBC.}
The next step is to investigate how sensitive estimated posterior model parameters are to the true simulating parameters for the \textcolor{black}{given} experimental design, statistical model, and set of prior parameters. We compute posterior z-scores to assess deviation of estimated means from true means, and compute posterior contraction to investigate how much information is gained in the posterior relative to the prior, that is, how much the uncertainty about a parameter of interest is reduced. We study this for the theoretically important effect size of object versus subject relative sentences (see Supplementary Code S12).

\begin{figure}

{\centering \includegraphics{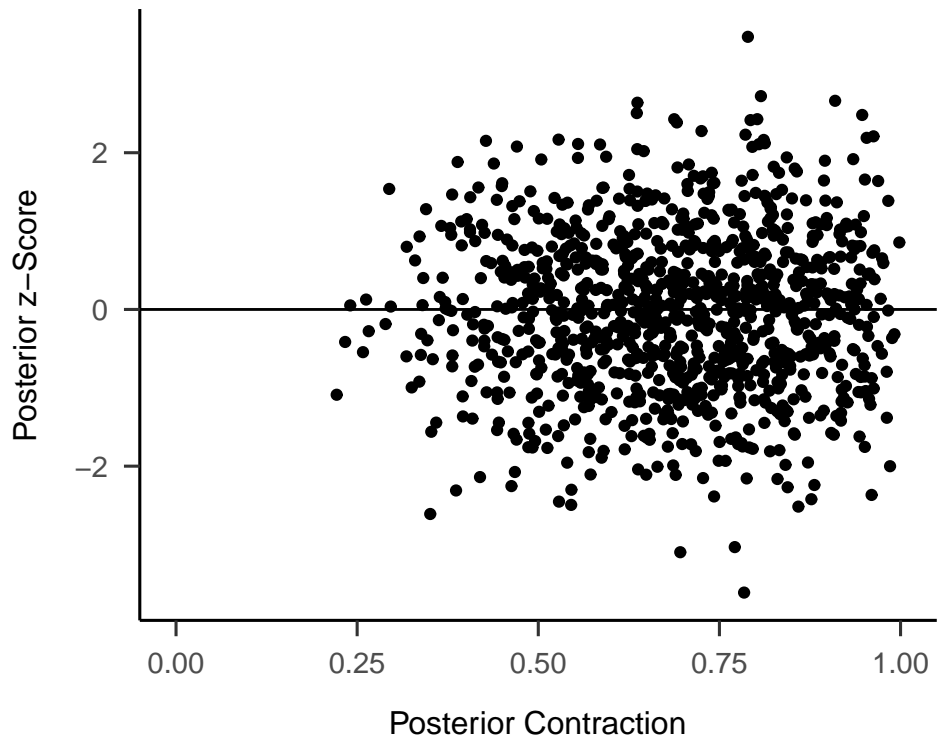} 

}

\caption{Analysis of model sensitivity. For each of N = 1,000 data-sets simulated from the prior parameters and the model, plot the posterior z-score as a function of the posterior contraction. The results show good posterior z-scores, as most of them are very close to zero, indicating little overfitting to wrong values or prior/likelihood conflict. The results for posterior contraction, however, are mixed. While some data-sets show very good (high) contraction close to one, others have only weak contraction of about 0.5, reflecting the relatively low number of participants and items in the experimental design.}\label{fig:FigSensitivity}
\end{figure}

\FloatBarrier

Figure~\ref{fig:FigSensitivity} shows posterior z-scores as a function of posterior contraction for all simulated data-sets. The results show that posterior z-scores are overall relatively close to zero, and mostly below absolute values of \(2\) (average: 0.82), reflecting good recovery of the ground truth. At the same time, posterior contraction ranges from only weak contraction (i.e., less than \(0.5\)), where not a lot of information is gained about the parameter in the posterior relative to the prior, to very strong contraction (i.e., approaching \(1\)), where posterior uncertainty is much reduced. The average contraction is 0.69.

The sometimes low posterior contraction indicates a tendency that for some simulated data-sets the slope parameter is not very well identified. This shortcoming may reflect the relatively small number of observed data points in the present study, reflecting \textcolor{black}{insufficient resolution to detect an effect}. Nevertheless, in \textcolor{black}{some of the} simulations model sensitivity looks reasonable.

This observation shows that posterior behavior will vary with the observed data, and that a model can be pathological for some data but not others. Since a priori we don't know what data we will ultimately observe, we want to check as many reasonable data-sets as possible before running the study, which is conveniently done in the prior predictive analyses.

\begin{figure}

{\centering \includegraphics{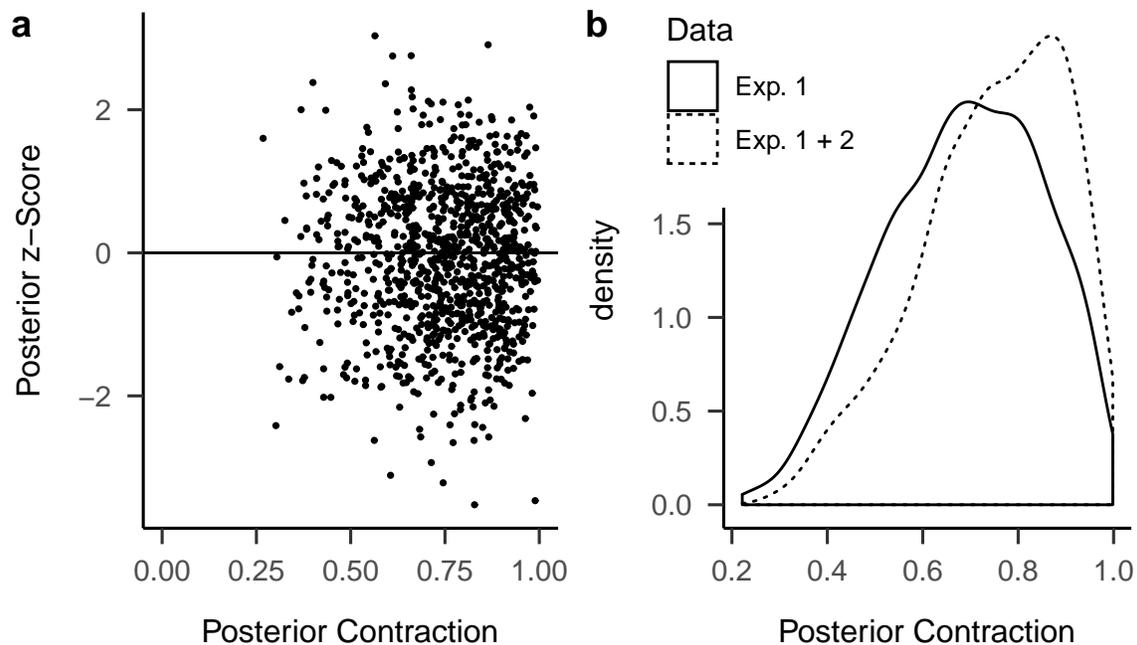} 

}

\caption{Analysis of model sensitivity for more observations, achieved by pooling across two data-sets. a) Results for pooled data-set (Exp 1: Gibson and Wu, 2012; Exp 2: Vasishth et al., 2013). b) Results only for posterior contraction for Experiment 1 (Gibson and Wu, 2012) and for the pooled data-set (Exp. 1 + 2). a)+b) The results show that - as the data now provides more information via the likelihood - this increased posterior contraction leads to more concentrated and hence more informative posteriors. a) At the same time, posterior z-scores are still rather close to zero, indicating good sensitivity.}\label{fig:FigSensitivityGW12}
\end{figure}

How does model sensitivity look like in a situation with \textcolor{black}{a higher resolution to detect an effect}? A replication study (Vasishth et al., 2013) is available for the present data-set by Gibson \& Wu (2013). Here, we combine the data from both studies to see how the associated increase in statistical \textcolor{black}{resolution} affects model sensitivity. The results show that posterior z-scores are relatively unchanged compared to the previous analysis, and mainly are between absolute values of \(2\) (average: 0.83). For the posterior contraction, however, the samples now cluster slightly more to the right of Figure~\ref{fig:FigSensitivityGW12}a and b: mean contraction is now 0.75 compared to the previous 0.69. This indicates somewhat stronger posterior contraction as a result of the higher number of subjects. On average, the data thus now provide more information on the parameter, and lead to a stronger reduction of uncertainty about the parameter of interest. At the same time, however, the amount of contraction strongly varies across different simulated data-sets. Even with the larger number of subjects, contraction can be quite low (i.e., values around \(0.5\)) for some simulated data-sets, whereas it is high for others. This means that even with double the number of subjects, we have no guarantee of getting informative results from our experiment.

\FloatBarrier

\hypertarget{posterior-predictive-checks-model-adequacy}{%
\subsection{Posterior predictive checks: Model adequacy}\label{posterior-predictive-checks-model-adequacy}}

Having examined the prior predictive data in detail, we can now take the real, observed data and perform posterior inference on it. We start by fitting a maximal \emph{brm} model to the observed data (see Supplementary Code S13).

\begin{Shaded}
\begin{Highlighting}[]
\NormalTok{m_gw <-}\StringTok{ }\KeywordTok{brm}\NormalTok{(rt }\OperatorTok{~}\StringTok{ }\NormalTok{so }\OperatorTok{+}\StringTok{ }\NormalTok{(}\DecValTok{1}\OperatorTok{+}\NormalTok{so}\OperatorTok{|}\NormalTok{subj) }\OperatorTok{+}\StringTok{ }\NormalTok{(}\DecValTok{1}\OperatorTok{+}\NormalTok{so}\OperatorTok{|}\NormalTok{item), gw1, }
            \DataTypeTok{family=}\KeywordTok{lognormal}\NormalTok{(), }\DataTypeTok{prior=}\NormalTok{priors2, }\DataTypeTok{cores=}\DecValTok{4}\NormalTok{)}
\end{Highlighting}
\end{Shaded}

\begin{figure}

{\centering \includegraphics{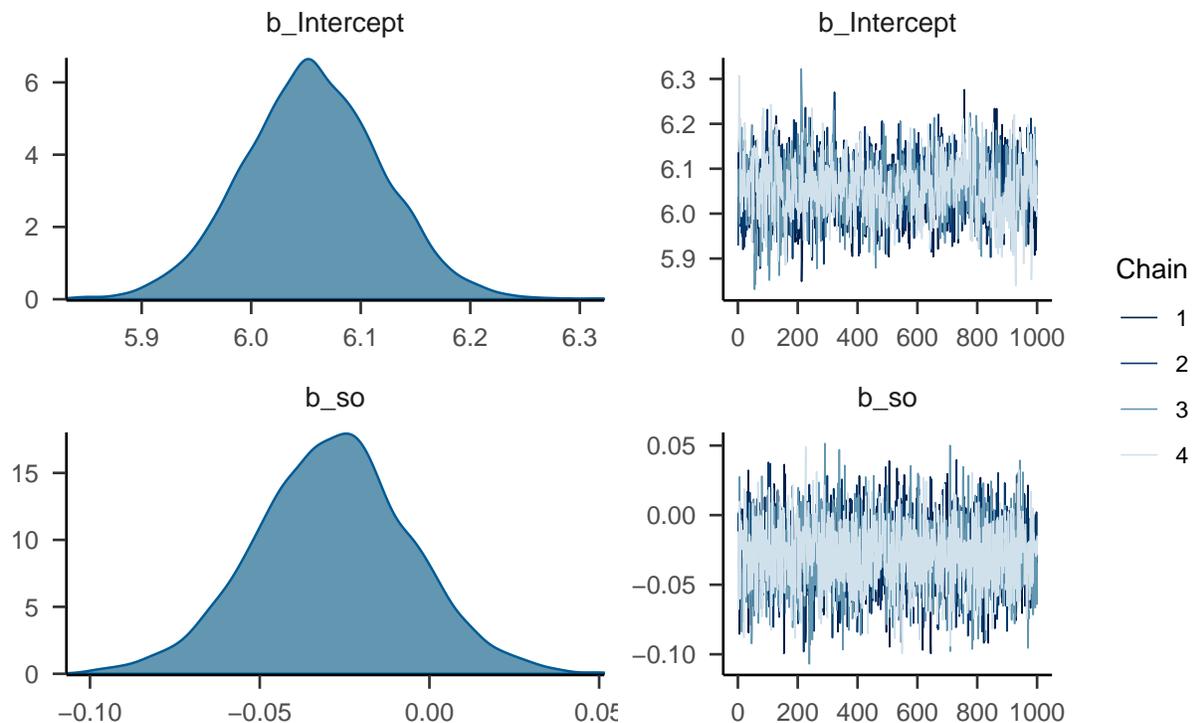} 

}

\caption{Density plots (left panels) and trace plots (right panels) for the intercept parameter (upper panels) and for the difference between subject and object relative sentences, labeled 'b so' (lower panels).}\label{fig:TracePlotsFinal}
\end{figure}

\textcolor{black}{We check visually whether the chains seem to have converged and whether they mix well} (see Fig.~\ref{fig:TracePlotsFinal}).

\FloatBarrier

\textcolor{black}{Next, we look at the estimated parameter values for the fixed effects:}

\begin{Shaded}
\begin{Highlighting}[]
\KeywordTok{round}\NormalTok{(}\KeywordTok{fixef}\NormalTok{(m_gw),}\DecValTok{3}\NormalTok{)}
\end{Highlighting}
\end{Shaded}

\begin{verbatim}
##           Estimate Est.Error   Q2.5 Q97.5
## Intercept    6.056     0.063  5.932 6.180
## so          -0.028     0.023 -0.075 0.017
\end{verbatim}

\begin{figure}

{\centering \includegraphics{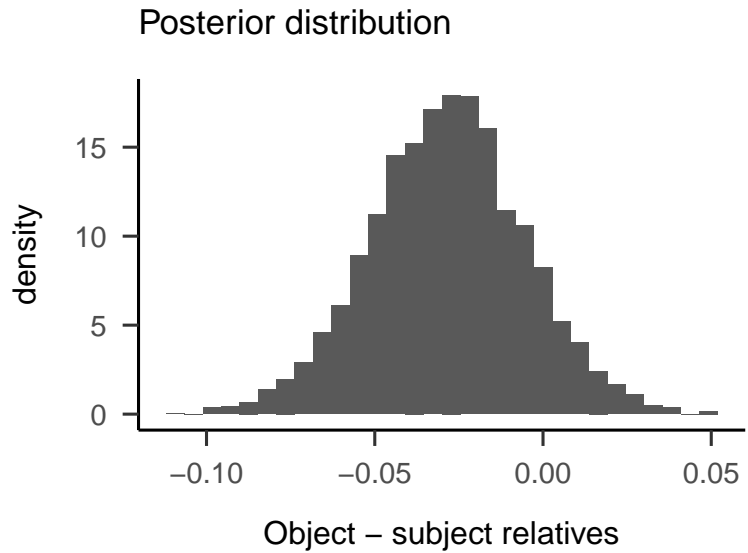} 

}

\caption{Posterior distribution for the slope parameter, estimating the difference in reading times between object relative minus subject relative sentences.}\label{fig:FigPosteriorB}
\end{figure}

Figure~\ref{fig:FigPosteriorB} shows the posterior distribution for the slope parameter, which estimates the difference in reading times between object minus subject relative sentences (see Supplementary Code S14).

\FloatBarrier

Figure~\ref{fig:FigPosteriorB} shows that the reading times in object relative sentences tends to be slightly faster than in subject relative sentences (p(b\textless{}0) = 0.89); this is as predicted by Gibson \& Wu (2013). The \(95\)\% confidence intervals, however, overlap with zero; it is difficult to rule out the possibility that there is effectively no difference in reading time between the two conditions. \textcolor{black}{Note that this analysis does not allow conclusions about whether no difference in reading times between conditions exist, since we have not included a model without any difference. We will do so below by using Bayes factor analyses.}

To assess model adequacy, we perform posterior predictive checks. We simulate data based on posterior samples of parameters. This then allows us to investigate the simulated data by computing the summary statistics that we used in the prior predictive checks, and by comparing model predictions with the observed data (see Supplementary Code S15).

\begin{figure}

{\centering \includegraphics{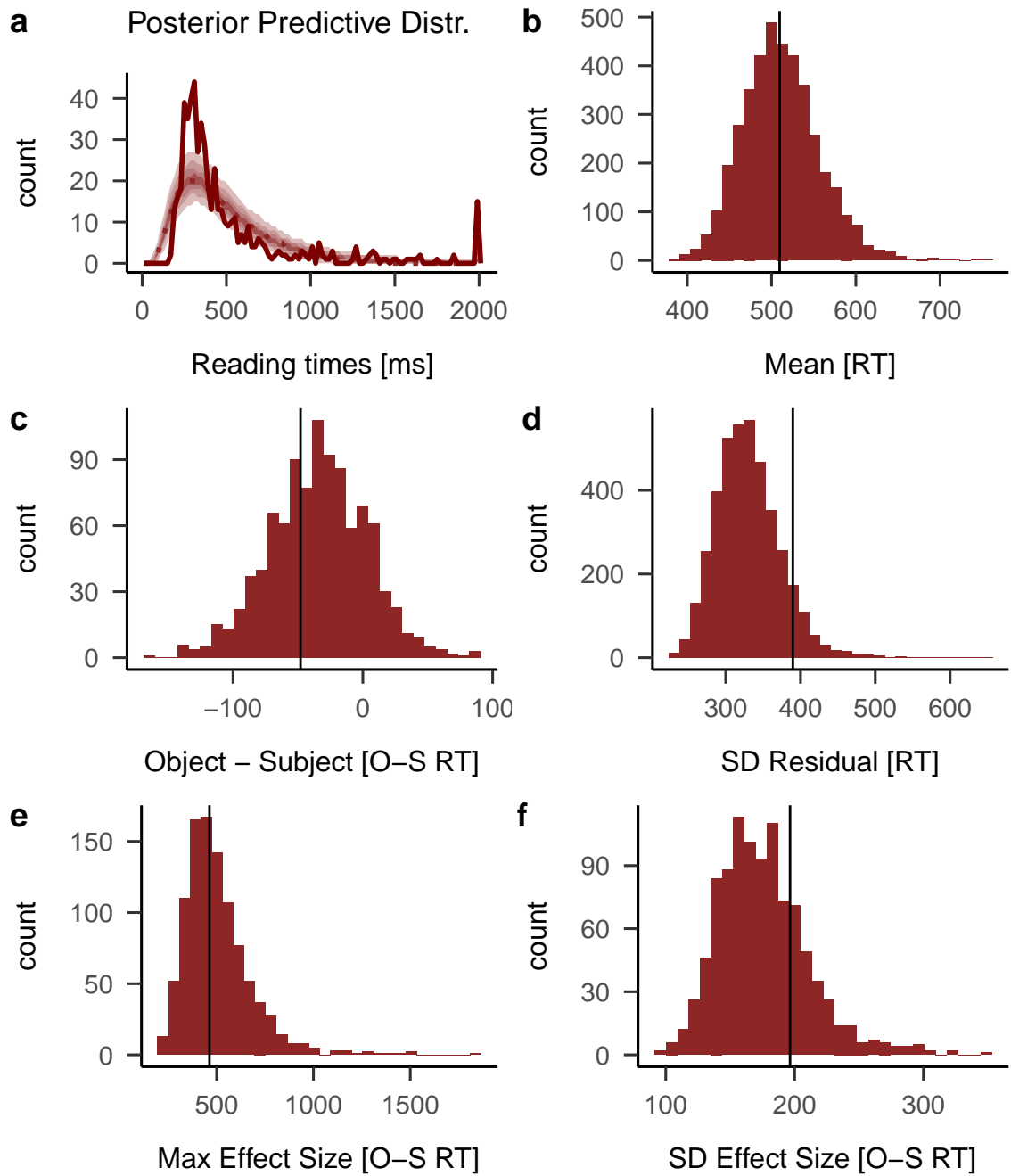} 

}

\caption{Posterior predictive checks for weakly informative priors. Distributions are over posterior predictive simulated data. a) Histograms of reading times. 10-90, 20-80, 30-70, and 40-60 percent quantiles across histograms are shown as shaded areas; the median is shown as a dotted line and the observed data as a solid line. For illustration, values > 2000 are plotted as 2000; modeling was done on the original data. b) Average reading times. c) Differences in reading times between object minus subject relatives. d) Standard deviations of residual reading times. e) Maximal effect size (object - subject relatives) across subjects. f) Standard deviation of effect size (object - subject relatives) across subjects.}\label{fig:FigPost}
\end{figure}

\FloatBarrier

These analyses show that the lognormal distribution (see Fig.~\ref{fig:FigPost}a) provides an approximation to the distribution of the data. Although the fit looks reasonable, however, there is still systematic deviation from the data of the model's predictions. This deviation suggests that maybe a constant offset is needed in addition to the lognormal distribution. This can be implemented in \emph{brm} by replacing the family specification \texttt{family=lognormal()} with the shifted version \texttt{family=shifted\_lognormal()}, and \textcolor{black}{would} motivate another round of model validation.

\textcolor{black}{For the other summary statistics, we first} look at the distribution of means. The posterior predictive means capture the mean reading time in the observed data (\textcolor{black}{indicated by} the vertical line in Fig.~\ref{fig:FigPost}b) \textcolor{black}{quite well}. The same is true for the \textcolor{black}{residual} standard deviation - the model captures the standard deviation of the data (Fig.~\ref{fig:FigPost}d). Figure~\ref{fig:FigPost}c shows the effect size of object minus subject relative sentences predicted by model (histogram) and observed in the data (vertical line). Again, posterior model predictions for the effect are in line with the empirical data. The same is true for the biggest effect among all subjects (Fig.~\ref{fig:FigPost}e) and for the \textcolor{black}{random effects} standard deviation of the effect across subjects (Fig.~\ref{fig:FigPost}f).

We had mentioned above that some subjects were removed due to invalid data. Note that in the frameworks of lme4 and brms these missing subjects cannot be modelled without adding new functionality to these packages. This, however, is possible when using Stan directly.

\hypertarget{bayes-factor-analysis}{%
\subsection{Bayes factor analysis}\label{bayes-factor-analysis}}

The posterior predictive checks suggest that the maximal model captures the summary statistics of the data well\textcolor{black}{, increasing our confidence} that we can rely on the model for interpreting the estimate for the effect size of \textcolor{black}{object minus subject} relatives (\enquote{so}). Simply testing the effect size, however, does not provide evidence on whether the effect of \textcolor{black}{relative clause type} exists, i.e., whether it is different from zero. To answer this question, we can compute Bayes factors, where we compare the maximal model to a reduced model, where the fixed effect of \textcolor{black}{the predictor} \enquote{so} is missing (which essentially sets its parameter to zero).

\textcolor{black}{Bayes factors provide a way to quantify the evidence that some data provide in favor of one model over another model. It evaluates the model and the prior by evaluating its prior predictive accuracy. More specifically, the (marginal) likelihood of the observed data given the model (including its priors) is computed for two models, and the ratio of (marginal) likelihoods is the Bayes factor:}

\begin{equation}
BF_{01} = \frac{p(D \mid \mathcal{M}_0)}{p(D \mid \mathcal{M}_1)} = 
\frac{
\int p(D\mid \boldsymbol{\theta}, \mathcal{M}_0) p(\boldsymbol{\theta} \mid \mathcal{M}_0) d\boldsymbol{\theta}
}{
\int p(D\mid \boldsymbol{\theta}, \mathcal{M}_1) p(\boldsymbol{\theta} \mid \mathcal{M}_1) d\boldsymbol{\theta}
}
\end{equation}

\textcolor{black}{For more detailed introductions to and treatments of Bayes factors, we refer to} Ly, Verhagen, \& Wagenmakers (2016), Mulder \& Wagenmakers (2016), Rouder, Haaf, \& Vandekerckhove (2018), and Schönbrodt \& Wagenmakers (2018). Also see Gronau et al. (2017) \textcolor{black}{for a tutorial on Bridge sampling, which we use to compute Bayes factors.}

\textcolor{black}{A very important point here is that in}
the Bayes factor analysis, the more informative priors that \textcolor{black}{were} derived from the prior predictive checks will be used.
To compute a Bayes factor in brm, a very large number of posterior samples are needed in order to obtain stable Bayes factor \textcolor{black}{values}. The model \textcolor{black}{is therefore re-fit} twice with larger number of samples (\texttt{iter=10000}), once with the fixed effect \enquote{so} included, and once without the fixed effect for \enquote{so} (see Supplementary Code S16).

\begin{Shaded}
\begin{Highlighting}[]
\NormalTok{m_gw1 <-}\StringTok{ }\KeywordTok{brm}\NormalTok{(rt }\OperatorTok{~}\StringTok{ }\NormalTok{so }\OperatorTok{+}\StringTok{ }\NormalTok{(}\DecValTok{1}\OperatorTok{+}\NormalTok{so}\OperatorTok{|}\NormalTok{subj) }\OperatorTok{+}\StringTok{ }\NormalTok{(}\DecValTok{1}\OperatorTok{+}\NormalTok{so}\OperatorTok{|}\NormalTok{item), gw1, }
            \DataTypeTok{family=}\KeywordTok{lognormal}\NormalTok{(), }\DataTypeTok{prior=}\NormalTok{priors2, }\DataTypeTok{cores=}\DecValTok{4}\NormalTok{, }
            \DataTypeTok{save_all_pars=}\OtherTok{TRUE}\NormalTok{, }\DataTypeTok{iter=}\DecValTok{10000}\NormalTok{, }\DataTypeTok{warmup=}\DecValTok{2000}\NormalTok{)}
\NormalTok{m_gw0 <-}\StringTok{ }\KeywordTok{brm}\NormalTok{(rt }\OperatorTok{~}\StringTok{ }\DecValTok{1} \OperatorTok{+}\StringTok{ }\NormalTok{(}\DecValTok{1}\OperatorTok{+}\NormalTok{so}\OperatorTok{|}\NormalTok{subj) }\OperatorTok{+}\StringTok{ }\NormalTok{(}\DecValTok{1}\OperatorTok{+}\NormalTok{so}\OperatorTok{|}\NormalTok{item), gw1, }
            \DataTypeTok{family=}\KeywordTok{lognormal}\NormalTok{(), }\DataTypeTok{prior=}\NormalTok{priors2[}\OperatorTok{-}\DecValTok{2}\NormalTok{,], }\DataTypeTok{cores=}\DecValTok{4}\NormalTok{, }
            \DataTypeTok{save_all_pars=}\OtherTok{TRUE}\NormalTok{, }\DataTypeTok{iter=}\DecValTok{10000}\NormalTok{, }\DataTypeTok{warmup=}\DecValTok{2000}\NormalTok{)}
\NormalTok{BF_informative <-}\StringTok{ }\KeywordTok{bayes_factor}\NormalTok{(m_gw1, m_gw0)}
\end{Highlighting}
\end{Shaded}

\begin{Shaded}
\begin{Highlighting}[]
\NormalTok{BF_informative}
\end{Highlighting}
\end{Shaded}

\begin{verbatim}
## Estimated Bayes factor in favor of bridge1 over bridge2: 0.65536
\end{verbatim}

The results show a Bayes factor \textcolor{black}{in favor of the model H1} of approximately \(0.65\)\textcolor{black}{, which translates into a Bayes factor in favor of H0 of $1.5$. This indicates that there is slight support for the null model over the maximal model, but that the data do not allow us to prefer any of the two models over the other.} Thus, it is not clear from this data-set whether there is a difference in reading times between Chinese subject and object relative clauses. \textcolor{black}{By contrast, the published study} (Gibson \& Wu, 2013) \textcolor{black}{reported a significant effect of subject versus object relative clauses (based on a repeated measures ANOVA) and concluded that the effect was present.}

The choice of informative priors is crucial for a valid analysis of Bayes factors (Ly et al., 2016; Mulder \& Wagenmakers, 2016; Rouder et al., 2018; Schönbrodt \& Wagenmakers, 2018). If the prior assumes that extremely large values for the \enquote{so} effect are possible, then the Bayes factor assesses whether there is evidence for such extremely large effect sizes. Of course, this is very unlikely to be the case empirically. \textcolor{black}{By contrast}, when using weakly informative priors (e.g., \textcolor{black}{those} informed by the prior predictive checks outlined above), then the prior assumes small to medium effect sizes for the \enquote{so} effect, and the Bayes factor accordingly tests whether there is evidence for such small or medium effect sizes.

Indeed, when we re-compute the Bayes factor between the maximal model and the reduced model, but using the vague or diffuse priors discussed above, the Bayes factor \textcolor{black}{(in support of H1)} is strongly reduced:

\begin{verbatim}
## Estimated Bayes factor in favor of bridge1 over bridge2: 0.03981
\end{verbatim}

\textcolor{black}{We now have strong support for the null model.}
The meaning of the Bayes factor is determined by the order in which the models are entered into the calculation. A Bayes factor of roughly\footnote{The \texttt{bayes\_factor()} command should be run several times to make sure that the number is stable with respect to the randomness inherent to bridge sampling estimators. \textcolor{black}{We are currently writing a paper discussing the instability of Bayes factor computations using Bridge sampling.}} \(0.04\) indicates that the first of the two models, here the \textcolor{black}{null} model, \textcolor{black}{receives more evidence from the data} than the second model, here the maximal model. Now, there is suddenly strong evidence for the null hypothesis! Note that the only thing that was changed was the prior. With the diffuse/vague priors employed in the second Bayes factor analysis, there is clear evidence that extremely large differences between subject and object relatives are not supported by the data.

These results on how Bayes factors are highly sensitive to the prior highlight the importance of using reasonable priors when \textcolor{black}{comparing} Bayesian models. Simply using diffuse/vague priors can strongly bias Bayes factors towards the reduced model. It is therefore crucial to use domain knowledge to encode reasonable expectations about the size of the expected effect(s) of interest into (weakly) informative priors.
A good strategy is to display a range of Bayes factors using increasingly informative priors. For an example, see Nicenboim, Vasishth, \& Rösler (2019).

\textcolor{black}{In this context, we note that it is possible to define critical parameters that are added, constrained, or removed in model expansion/deflation, and nuisance parameters that are needed in all models. Examples of nuisance parameters are a grand mean intercept parameter and a residual variance parameter.} \textcolor{black}{These parameters do not define the differences among models between the initial model and the aspirational model. Thus, uninformative priors may be adequate for nuisance parameters when computing Bayes factors. Importantly, however, nuisance parameters can still have a strong influence on model comparisons using Bayes factor. For example, in linear models a strong prior on measurement variability (i.e., the residual variance) allows weaker data to differentiate better between models and drastically affects the Bayes factor. Moreover,} \textcolor{black}{in a log-normal distribution as assumed in our example analysis, the prior on the intercept parameter also has an influence on what we expect for the effect size. Therefore, the nuisance grand mean intercept is not completely independent of the critical effect size effect. We therefore (and for reasons of sampling speed, see below) recommend using reasonably constrained priors even on nuisance parameters.}

\textcolor{black}{One of the biggest motivations of a principled workflow is to consider a model in its entirety so that we don’t have to worry about what terms cancel or how influences factorize. We just analyze the entire model and all of its joint consequences.}

An important open question we do not discuss in the present paper is that the Bayes factor itself will be quite variable under repeated sampling; the stability of the Bayes factor will be discussed in a future paper.

An additional benefit of incorporating more domain knowledge into the prior is that this speeds up posterior sampling. For example, consider the maximal model for the \enquote{so} effect: for this, we recorded the time it took (in seconds) to fit one model and to perform the Bayes factor analysis (using the command \texttt{proc.time()}), and we did so for the diffuse priors and for the weakly informative priors (see Supplementary Code S17).

Incorporating domain knowledge into the priors lead to a speed-up in fitting the brm model and running the Bayes factor analysis from 152 seconds for the diffuse/vague priors to 102 seconds for the weakly informative priors. With more complex models, larger data sets, or when investigating computational faithfulness or model sensitivity, these time differences can become substantial.

\textcolor{black}{Note that we here present one possible perspective on treating priors in Bayesian modeling. In this exposition, the prior does not carry strong theoretical weight. We aim for inference on model parameters. In Bayesian data analysis, the priors are necessary even though we may not have very strong theoretical a priori constraints for the model parameters. In the principled Bayesian workflow, the goal therefore is to understand the role of the prior for the parameters, and to} \textcolor{black}{and to ensure that the influence of the priors is consistent with ones principled domain expertise.}

\textcolor{black}{For completeness, note that a different perspective on the role of priors in Bayesian analyses is also possible. In this alternative perspective or approach, the prior defines theoretically useful properties, and it is chosen not based on substantive ranges but on theoretical constraints} (Vandekerckhove et al., 2018). \textcolor{black}{In this case, because the prior is part of the model, methods for expanding or reducing the models should be sensitive to the model. Importantly, this includes being sensitive to the prior. Therefore, the sensitivity of the Bayes factor to the prior provides a very useful aspect of Bayesian modeling that allows the researcher to test substantive cognitive questions by constraining prior assumptions about the parameters to compare different models. An aspect of this can be implemented in sensitivity analyses, where different assumptions about the priors are entertained, and Bayes factors are used to compare models which differ only in their priors, in order to learn about which priors are best supported by the data.}

\hypertarget{summary}{%
\section{Summary}\label{summary}}

We have introduced key questions to ask about a model and the inference process as discussed by Betancourt (2018), and have applied this to a data-set from an experiment involving a typical repeated measures experimental design used in cognitive psychology and psycholinguistics. Prior predictive checks using analyses of simulated prior data suggest that, compared to previous applications in reading experiments (e.g., Nicenboim \& Vasishth, 2018), far more informative priors can and should be used. We \textcolor{black}{showed} that including such additional domain knowledge into the priors leads to more plausible expected data. Moreover, incorporating more informative priors \textcolor{black}{(if they change the posterior) can} also speed up the HMC sampling process. These more informative priors, however, may not alter posterior inferences much for the present design. We also investigated computational faithfulness using simulation-based calibration (SBC) and \textcolor{black}{showed} that prior model parameters were well recovered using posterior estimation, supporting the used estimation procedure (\texttt{brm()} function as a wrapper to Stan) for the current setup. Analysis of model sensitivity showed that the critical theoretical effect of a psycholinguistic manipulation was estimated without bias as posterior z-scores were centered around zero. Posterior contraction varied between medium and strong contraction, indicating somewhat weak \textcolor{black}{statistical sensitivity} in a rather small sample size. In line with this limited model sensitivity, posterior inference on the experimental effect based on the observed data did not provide strong evidence for the experimental effect of interest, leaving uncertain whether it differs from zero. Posterior predictive checks showed strong support for our statistical model, as the model successfully recovered most of the tested summary statistics.
The Bayes factor analysis showed \textcolor{black}{some weak} evidence for no effect (with diffuse priors) and an inconclusive result (with informative priors). Our overall conclusion would be that we did not learn much from the experiment. As an aside, note that the published result in Gibson \& Wu (2013) showed a statistically significant effect (using repeated measures ANOVA) and concluded that the effect was present. \textcolor{black}{Although this significant effect was due to model misspecification} (Vasishth et al., 2013)\textcolor{black}{, in general it can happen that the results of a frequentist and Bayesian analysis do not yield the same conclusion. This is an instance of what is sometimes termed "Lindley's paradox"} (Lindley, 1957), \textcolor{black}{which describes situations where the results from Bayesian and frequentist hypothesis tests differ from each other.} \textcolor{black}{Importantly, however, frequentist null hypothesis significance testing and a Bayesian decision making process are different things, and a certain calibration of one does not imply the same calibration of the other. Therefore, the differences that we see between them are expected, and not really a paradox.}

In summary, this analysis provides a fully worked example and tutorial for using the principled Bayesian workflow (Betancourt, 2018) in cognitive science experiments. The workflow reveals useful information about which (weakly informative) priors to use, and performs checks of the used inference procedures and the statistical model. The workflow provides a robust foundation for using a statistical model to answer scientific questions, and will be useful for researchers developing analysis plans as part of preregistrations, registered reports, or simply as preparatory design analyses prior to conducting an experiment.

\hypertarget{acknowledgements}{%
\section{Acknowledgements}\label{acknowledgements}}

Thanks to Bruno Nicenboim, Titus von der Malsburg, and participants of a lab meeting of the Vasishth lab for discussion. This work was partly funded by the Deutsche Forschungsgemeinschaft (DFG, German Research Foundation) -- project number 317633480 -- SFB 1287, Project Q (Principal Investigators Shravan Vasishth and Ralf Engbert).

\newpage

\hypertarget{references}{%
\section{References}\label{references}}

\begingroup
\setlength{\parindent}{-0.5in}
\setlength{\leftskip}{0.5in}

\hypertarget{refs}{}
\leavevmode\hypertarget{ref-baayen2008mixed}{}%
Baayen, R. H., Davidson, D. J., \& Bates, D. M. (2008). Mixed-effects modeling with crossed random effects for subjects and items. \emph{Journal of Memory and Language}, \emph{59}(4), 390--412.

\leavevmode\hypertarget{ref-Barr:2013aa}{}%
Barr, D. J., Levy, R., Scheepers, C., \& Tily, H. J. (2013). Random effects structure for confirmatory hypothesis testing: Keep it maximal. \emph{Journal of Memory and Language}, \emph{68}(3), 255--278.

\leavevmode\hypertarget{ref-lme4}{}%
Bates, D. M., Mächler, M., Bolker, B., \& Walker, S. (2015). Fitting linear mixed-effects models using lme4. \emph{Journal of Statistical Software}, \emph{67}(1), 1--48. \url{https://doi.org/10.18637/jss.v067.i01}

\leavevmode\hypertarget{ref-Betancourt:2018aa}{}%
Betancourt, M. (2018). Towards a principled Bayesian workflow. Retrieved from \url{https://betanalpha.github.io/assets/case_studies/principled_bayesian_workflow.html}

\leavevmode\hypertarget{ref-box1979robustness}{}%
Box, G. E. (1979). Robustness in the strategy of scientific model building. In \emph{Robustness in statistics} (pp. 201--236). Elsevier.

\leavevmode\hypertarget{ref-R-brms_b}{}%
Bürkner, P.-C. (2017a). Advanced Bayesian multilevel modeling with the R package brms. \emph{arXiv Preprint arXiv:1705.11123}.

\leavevmode\hypertarget{ref-Buerkner:2017aa}{}%
Bürkner, P.-C. (2017b). brms: An R package for Bayesian multilevel models using Stan. \emph{Journal of Statistical Software}, \emph{80}(1), 1--28.

\leavevmode\hypertarget{ref-carpenter2017stan}{}%
Carpenter, B., Gelman, A., Hoffman, M. D., Lee, D., Goodrich, B., Betancourt, M., \ldots{} Riddell, A. (2017). Stan: A probabilistic programming language. \emph{Journal of Statistical Software}, \emph{76}(1).

\leavevmode\hypertarget{ref-chow2017bayesian}{}%
Chow, S.-M., \& Hoijtink, H. (2017). Bayesian estimation and modeling: Editorial to the second special issue on Bayesian data analysis. \emph{Psychological Methods}, \emph{22}(4), 609--615.

\leavevmode\hypertarget{ref-cook2006validation}{}%
Cook, S. R., Gelman, A., \& Rubin, D. B. (2006). Validation of software for Bayesian models using posterior quantiles. \emph{Journal of Computational and Graphical Statistics}, \emph{15}(3), 675--692.

\leavevmode\hypertarget{ref-etz2018how}{}%
Etz, A., Gronau, Q. F., Dablander, F., Edelsbrunner, P. A., \& Baribault, B. (2018). How to become a bayesian in eight easy steps: An annotated reading list. \emph{Psychonomic Bulletin \& Review}, \emph{25}(1), 219--234.

\leavevmode\hypertarget{ref-etz2018introduction}{}%
Etz, A., \& Vandekerckhove, J. (2018). Introduction to bayesian inference for psychology. \emph{Psychonomic Bulletin \& Review}, \emph{25}(1), 5--34.

\leavevmode\hypertarget{ref-Gabry:2017aa}{}%
Gabry, J., Simpson, D., Vehtari, A., Betancourt, M., \& Gelman, A. (2017). Visualization in Bayesian workflow. \emph{arXiv Preprint arXiv:1709.01449}.

\leavevmode\hypertarget{ref-Gelman14}{}%
Gelman, A., Carlin, J. B., Stern, H. S., Dunson, D. B., Vehtari, A., \& Rubin, D. B. (2014). \emph{Bayesian data analysis} (Third). Boca Raton, FL: Chapman; Hall/CRC.

\leavevmode\hypertarget{ref-gelman2017prior}{}%
Gelman, A., Simpson, D., \& Betancourt, M. (2017). The prior can often only be understood in the context of the likelihood. \emph{Entropy}, \emph{19}(10), 555.

\leavevmode\hypertarget{ref-Gibson:2013aa}{}%
Gibson, E., \& Wu, H.-H. I. (2013). Processing Chinese relative clauses in context. \emph{Language and Cognitive Processes}, \emph{28}(1-2), 125--155.

\leavevmode\hypertarget{ref-Good:1950aa}{}%
Good, I. J. (1950). \emph{Probability and the weighing of evidence}. New York: Hafners.

\leavevmode\hypertarget{ref-goodrich2018rstanarm}{}%
Goodrich, B., Gabry, J., Ali, I., \& Brilleman, S. (2018). Rstanarm: Bayesian applied regression modeling via stan. \emph{R Package Version 2.17.4. Http://Mc-Stan.org/}, \emph{2}(4).

\leavevmode\hypertarget{ref-gronau2017tutorial}{}%
Gronau, Q. F., Sarafoglou, A., Matzke, D., Ly, A., Boehm, U., Marsman, M., \ldots{} Steingroever, H. (2017). A tutorial on bridge sampling. \emph{Journal of Mathematical Psychology}, \emph{81}, 80--97.

\leavevmode\hypertarget{ref-haaf2017developing}{}%
Haaf, J. M., \& Rouder, J. N. (2017). Developing constraint in bayesian mixed models. \emph{Psychological Methods}, \emph{22}(4), 779.

\leavevmode\hypertarget{ref-hoijtink2017bayesian}{}%
Hoijtink, H., \& Chow, S.-M. (2017). Bayesian hypothesis testing: Editorial to the special issue on Bayesian data analysis. \emph{Psychological Methods}, \emph{22}(2), 211--216.

\leavevmode\hypertarget{ref-JASP2019}{}%
JASP Team. (2019). JASP (Version 0.11.1){[}Computer software{]}. Retrieved from \url{https://jasp-stats.org/}

\leavevmode\hypertarget{ref-kmenta1971elements}{}%
Kmenta, J. (1971). \emph{Elements of econometrics}. New York: Macmillan; London: Collier Macmillan.

\leavevmode\hypertarget{ref-kruschke2014doing}{}%
Kruschke, J. (2014). \emph{Doing bayesian data analysis: A tutorial with r, jags, and stan}. Academic Press.

\leavevmode\hypertarget{ref-lambert2018student}{}%
Lambert, B. (2018). \emph{A student's guide to Bayesian statistics}. Sage.

\leavevmode\hypertarget{ref-lee2011cognitive}{}%
Lee, M. D. (2011). How cognitive modeling can benefit from hierarchical Bayesian models. \emph{Journal of Mathematical Psychology}, \emph{55}(1), 1--7.

\leavevmode\hypertarget{ref-lee2019robust}{}%
Lee, M. D., Criss, A. H., Devezer, B., Donkin, C., Etz, A., Leite, F. P., \ldots{} others. (2019). Robust modeling in cognitive science. \emph{Computational Brain \& Behavior}, \emph{2}(3-4), 141--153.

\leavevmode\hypertarget{ref-lee2014bayesian}{}%
Lee, M. D., \& Wagenmakers, E.-J. (2014). \emph{Bayesian cognitive modeling: A practical course}. Cambridge University Press.

\leavevmode\hypertarget{ref-Lewandowski:2009aa}{}%
Lewandowski, D., Kurowicka, D., \& Joe, H. (2009). Generating random correlation matrices based on vines and extended onion method. \emph{Journal of Multivariate Analysis}, \emph{100}(9), 1989--2001.

\leavevmode\hypertarget{ref-lindley1957statistical}{}%
Lindley, D. V. (1957). A statistical paradox. \emph{Biometrika}, \emph{44}(1/2), 187--192.

\leavevmode\hypertarget{ref-link2012thinning}{}%
Link, W. A., \& Eaton, M. J. (2012). On thinning of chains in MCMC. \emph{Methods in Ecology and Evolution}, \emph{3}(1), 112--115.

\leavevmode\hypertarget{ref-lunn2000winbugs}{}%
Lunn, D., Thomas, A., Best, N., \& Spiegelhalter, D. J. (2000). WinBUGS-A Bayesian modelling framework: Concepts, structure, and extensibility. \emph{Statistics and Computing}, \emph{10}(4), 325--337.

\leavevmode\hypertarget{ref-ly2017flexible}{}%
Ly, A., Boehm, U., Heathcote, A., Turner, B. M., Forstmann, B., Marsman, M., \& Matzke, D. (2017). A flexible and efficient hierarchical bayesian approach to the exploration of individual differences in cognitive-model-based neuroscience. \emph{Computational Models of Brain and Behavior}, 467--480.

\leavevmode\hypertarget{ref-ly2016harold}{}%
Ly, A., Verhagen, J., \& Wagenmakers, E.-J. (2016). Harold Jeffreys's default Bayes factor hypothesis tests: Explanation, extension, and application in psychology. \emph{Journal of Mathematical Psychology}, \emph{72}, 19--32.

\leavevmode\hypertarget{ref-MacKay2003information}{}%
MacKay, D. J. (2003). \emph{Information theory, inference and learning algorithms}. Cambridge University Press.

\leavevmode\hypertarget{ref-morey2011bayesian}{}%
Morey, R. D. (2011). A bayesian hierarchical model for the measurement of working memory capacity. \emph{Journal of Mathematical Psychology}, \emph{55}(1), 8--24.

\leavevmode\hypertarget{ref-morey2016philosophy}{}%
Morey, R. D., Romeijn, J.-W., \& Rouder, J. N. (2016). The philosophy of bayes factors and the quantification of statistical evidence. \emph{Journal of Mathematical Psychology}, \emph{72}, 6--18.

\leavevmode\hypertarget{ref-mulder2016editors}{}%
Mulder, J., \& Wagenmakers, E.-J. (2016). Editors' introduction to the special issue ``Bayes factors for testing hypotheses in psychological research: Practical relevance and new developments''. \emph{Journal of Mathematical Psychology}, \emph{72}, 1--5.

\leavevmode\hypertarget{ref-Nicenboim:2018aa}{}%
Nicenboim, B., \& Vasishth, S. (2018). Models of retrieval in sentence comprehension: A computational evaluation using Bayesian hierarchical modeling. \emph{Journal of Memory and Language}, \emph{99}, 1--34.

\leavevmode\hypertarget{ref-nicenboim2019words}{}%
Nicenboim, B., Vasishth, S., \& Rösler, F. (2019). Are words pre-activated probabilistically during sentence comprehension? Evidence from new data and a Bayesian random-effects meta-analysis using publicly available data. \emph{PsyArXiv}.

\leavevmode\hypertarget{ref-nilsson2011hierarchical}{}%
Nilsson, H., Rieskamp, J., \& Wagenmakers, E.-J. (2011). Hierarchical bayesian parameter estimation for cumulative prospect theory. \emph{Journal of Mathematical Psychology}, \emph{55}(1), 84--93.

\leavevmode\hypertarget{ref-Paape:2017aa}{}%
Paape, D., Nicenboim, B., \& Vasishth, S. (2017). Does antecedent complexity affect ellipsis processing? An empirical investigation. \emph{Glossa: A Journal of General Linguistics}, \emph{2}(1).

\leavevmode\hypertarget{ref-pinheiro2000linear}{}%
Pinheiro, J. C., \& Bates, D. M. (2000). Linear mixed-effects models: Basic concepts and examples. \emph{Mixed-Effects Models in S and S-Plus}, 3--56.

\leavevmode\hypertarget{ref-plummer2011jags}{}%
Plummer, M. (2012). JAGS version 3.3.0 manual. \emph{International Agency for Research on Cancer. Lyon, France}.

\leavevmode\hypertarget{ref-pooley2011understanding}{}%
Pooley, J. P., Lee, M. D., \& Shankle, W. R. (2011). Understanding memory impairment with memory models and hierarchical bayesian analysis. \emph{Journal of Mathematical Psychology}, \emph{55}(1), 47--56.

\leavevmode\hypertarget{ref-pratte2011hierarchical}{}%
Pratte, M. S., \& Rouder, J. N. (2011). Hierarchical single-and dual-process models of recognition memory. \emph{Journal of Mathematical Psychology}, \emph{55}(1), 36--46.

\leavevmode\hypertarget{ref-rabe2020hypr}{}%
Rabe, M. M., Vasishth, S., Hohenstein, S., Kliegl, R., \& Schad, D. J. (2020). Hypr: An r package for hypothesis-driven contrast coding. \emph{PsyArXiv}. Retrieved from \url{10.31234/osf.io/cqzdx}

\leavevmode\hypertarget{ref-ratcliff2000diffusion}{}%
Ratcliff, R., \& Rouder, J. N. (2000). A diffusion model account of masking in two-choice letter identification. \emph{Journal of Experimental Psychology: Human Perception and Performance}, \emph{26}(1), 127.

\leavevmode\hypertarget{ref-RCoreTeam:2016aa}{}%
R Core Team. (2016). \emph{R: A language and environment for statistical computing}. Vienna, Austria: R Foundation for Statistical Computing.

\leavevmode\hypertarget{ref-rouder2018bayesian}{}%
Rouder, J. N., Haaf, J. M., \& Vandekerckhove, J. (2018). Bayesian inference for psychology, Part IV: Parameter estimation and Bayes factors. \emph{Psychonomic Bulletin \& Review}, \emph{25}(1), 102--113.

\leavevmode\hypertarget{ref-schad2018capitalize}{}%
Schad, D. J., Hohenstein, S., Vasishth, S., \& Kliegl, R. (2020). How to capitalize on a priori contrasts in linear (mixed) models: A tutorial. \emph{Journal of Memory and Language}, \emph{110}, 104038.

\leavevmode\hypertarget{ref-schonbrodt2018bayes}{}%
Schönbrodt, F. D., \& Wagenmakers, E.-J. (2018). Bayes factor design analysis: Planning for compelling evidence. \emph{Psychonomic Bulletin \& Review}, \emph{25}(1), 128--142.

\leavevmode\hypertarget{ref-shiffrin2008survey}{}%
Shiffrin, R. M., Lee, M. D., Kim, W., \& Wagenmakers, E.-J. (2008). A survey of model evaluation approaches with a tutorial on hierarchical Bayesian methods. \emph{Cognitive Science}, \emph{32}(8), 1248--1284.

\leavevmode\hypertarget{ref-Spiegelhalter:2014aa}{}%
Spiegelhalter, D. J., Best, N. G., Carlin, B. P., \& Linde, A. (2014). The Deviance Information Criterion: 12 years on. \emph{Journal of the Royal Statistical Society: Series B (Statistical Methodology)}, \emph{76}(3), 485--493.

\leavevmode\hypertarget{ref-Talts:2018aa}{}%
Talts, S., Betancourt, M., Simpson, D., Vehtari, A., \& Gelman, A. (2018). Validating Bayesian inference algorithms with simulation-based calibration. \emph{arXiv Preprint arXiv:1804.06788}.

\leavevmode\hypertarget{ref-vandekerckhove2018bayesian}{}%
Vandekerckhove, J., Rouder, J. N., \& Kruschke, J. K. (2018). Bayesian methods for advancing psychological science. \emph{Psychonomic Bulletin \& Review}, \emph{25}(1), 1--4.

\leavevmode\hypertarget{ref-VasishthetalPLoSOne2013}{}%
Vasishth, S., Chen, Z., Li, Q., \& Guo, G. (2013). Processing Chinese relative clauses: Evidence for the subject-relative advantage. \emph{PLoS ONE}, \emph{8}(10), 1--14.

\leavevmode\hypertarget{ref-Vasishth:2018aa}{}%
Vasishth, S., Mertzen, D., Jäger, L. A., \& Gelman, A. (2018a). The statistical significance filter leads to overoptimistic expectations of replicability. \emph{Journal of Memory and Language}, \emph{103}, 151--175.

\leavevmode\hypertarget{ref-vasishth2018bayesian}{}%
Vasishth, S., Nicenboim, B., Beckman, M. E., Li, F., \& Kong, E. J. (2018b). Bayesian data analysis in the phonetic sciences: A tutorial introduction. \emph{Journal of Phonetics}, \emph{71}, 147--161.

\leavevmode\hypertarget{ref-Vehtari:2017aa}{}%
Vehtari, A., Gelman, A., \& Gabry, J. (2017). Practical Bayesian model evaluation using leave-one-out cross-validation and WAIC. \emph{Statistics and Computing}, \emph{27}(5), 1413--1432.

\leavevmode\hypertarget{ref-vehtari2019rank}{}%
Vehtari, A., Gelman, A., Simpson, D., Carpenter, B., \& Bürkner, P.-C. (2019). Rank-normalization, folding, and localization: An improved \(\widehat R\) for assessing convergence of MCMC. \emph{arXiv Preprint arXiv:1903.08008}.

\leavevmode\hypertarget{ref-wagenmakers2016bayesian}{}%
Wagenmakers, E.-J., Morey, R. D., \& Lee, M. D. (2016). Bayesian benefits for the pragmatic researcher. \emph{Current Directions in Psychological Science}, \emph{25}(3), 169--176.

\endgroup

\end{document}